\documentclass[aps,prc,groupedaddress,twocolumn,nofootinbib]{revtex4-1} %DIF > 
\usepackage{bm}
\usepackage{graphicx}
\usepackage{ascmac}
\usepackage{amsmath,amssymb}
\usepackage{cancel}
\usepackage{braket}
\usepackage{url}
\usepackage[setpagesize=false,breaklinks=true]{hyperref} % breaklinks needs the dvipdfmx driver stated explicitly %DIF > 
\usepackage{xurl} % allow URLs (e.g. in the bibliography) to break at any character, avoiding overfull lines %DIF > 
\usepackage{seqsplit} % breakable display text for very long URLs that xurl alone does not wrap %DIF > 
\usepackage{docmute}
\usepackage{calligra}
\usepackage{color}
\usepackage{physics}

\newcommand{\ff}{\text{I\hspace{-1.2pt}I\;}} % \; はスペース入れるために必要
\allowdisplaybreaks[1]

\begin{document}

%\preprint{}

\title{Isospin-breaking effects on the threshold cusp structures \\
in $\Lambda N$-$\Sigma N$ scattering}


\author{Katsuyoshi Sone}
\email[]{sone-katsuyoshi@ed.tmu.ac.jp}
\affiliation{Department of Physics, Tokyo Metropolitan University, Hachioji 192-0397, Japan}
\author{Yudai Ichikawa}
\email[]{yudai.ichikawa.d3@tohoku.ac.jp}
\affiliation{Department of Physics, Tohoku University, Sendai 980-8578, Japan}
\author{Tetsuo Hyodo}
\email[]{hyodo@rcnp.osaka-u.ac.jp}
\affiliation{Research Center for Nuclear Physics (RCNP), The University of Osaka, 10-1 Mihogaoka, Ibaraki, Osaka 567-0047, Japan}


\date{\today}

\begin{abstract}
    We discuss the isospin symmetry breaking effects on threshold cusp structures in multichannel scattering near two-body thresholds. In hadronic systems with isospin symmetry, two or more nearly degenerate thresholds can appear, and their small splitting due to isospin breaking can generate multiple cusp structures in a narrow energy region. In this paper, using the $K$-matrix representation, we derive a general expression for the scattering amplitude near the thresholds and show that the cusp structures can be classified by the signs of the slopes of the cross section above and below threshold. We also show that additional restrictions appear in two- or three-channel systems and in the Flatt\'e amplitude. For three-channel scattering with two nearby thresholds, we clarify how the two cusp structures are related when the threshold splitting is small and how they merge into a single cusp in the degenerate limit. Finally, we discuss the cusp structures in the $\Lambda p$ elastic cross section in the coupled $\Lambda p$-$\Sigma^{+}n$-$\Sigma^{0}p$ system. We show that, when isospin breaking is small, the properties of the two cusp structures are constrained by isospin symmetry. We also perform quantitative calculations using both simplified examples and realistic input based on N$^2$LO chiral effective field theory, and find that isospin breaking can significantly modify the relative sharpness of the cusps and may even change the cusp type itself.
\end{abstract}

%\keywords{}

\maketitle


%%%%%%%%%%%%%%%%%%%%%%%%%%%%%%%%%%%%%%%%
\section{Introduction}
%%%%%%%%%%%%%%%%%%%%%%%%%%%%%%%%%%%%%%%%

% cusp sctructure
Threshold cusp structures are characteristic nonanalytic phenomena in multichannel scattering~\cite{Wigner:1948zz,Badalian:1981xj,Guo:2019twa}. When a new channel opens, unitarity and channel coupling generate a cusp in the line shape of the cross section at the corresponding threshold, and the resulting structure often contains useful information on the underlying interaction. For this reason, cusp phenomena have long been studied in various contexts. In hadron physics, threshold cusps are of particular interest because they can mimic resonance-like enhancements and, at the same time, provide direct information on coupled-channel dynamics near thresholds~\cite{Baru:2003qq,Cabibbo:2004gq,Baru:2004xg,Hyodo:2011js,Guo:2019twa}.

% scattering length determination
A well-known application of cusp analysis is the determination of scattering lengths. The extraction of the $\pi\pi$ scattering lengths from weak decays of $K$ mesons into the $3\pi$ final state was theoretically proposed~\cite{Budini:1961bac,Meissner:1997fa,Cabibbo:2004gq,Cabibbo:2005ez} and subsequently implemented in experiments~\cite{Batley:2005ax,Batley:2009ubw}. The $\pi\pi$ scattering lengths were determined by analyzing the cusp structure at the $\pi^{+}\pi^{-}$ threshold in the $\pi^{0}\pi^{0}$ spectrum. 
Similar ideas, utilizing cusps induced by final-state interactions in weak decays, have been applied to the determination of the $\pi\Sigma$ scattering lengths in $\Lambda_{c}\to \pi\pi\Sigma$ decays~\cite{Hyodo:2011js} and the $DN$ scattering lengths in $\Lambda_{b}\to \pi DN$ decays~\cite{Sakai:2020psu}. In addition, the $\Sigma N$ cusp structures are theoretically discussed within the chiral EFT framework~\cite{Haidenbauer:2021smk}. Also, attempts have been proposed to extract the $\Sigma N$ scattering lengths from the $\Sigma N$ cusp in the $K^{-}d\to p\Lambda\pi^{-}$ reaction~\cite{Badalian:1981xj,Dalitz:1982tb}. The $\Sigma N$ cusp is experimentally observed in the inclusive missing mass spectrum of the $d(\pi^+,K^+)$ reaction at J-PARC~\cite{Ichikawa:2014rva}.

% exotic hadrons
The cusp structures also play an important role in the study of exotic hadrons~\cite{Guo:2017jvc,Brambilla:2019esw,Hanhart:2022qxq,Hanhart:2025bun}. Many exotic hadron candidates appear near thresholds, and their observed line shapes are often influenced not only by poles of the scattering amplitude but also by kinematic and coupled-channel effects associated with nearby thresholds. In particular, threshold cusps are enhanced when a resonance pole lies near the threshold, accompanied by a large scattering length. In such situations, threshold cusp structures must be understood quantitatively in order to distinguish genuine near-threshold states from threshold-induced enhancements and to clarify the dynamics of channel coupling. Thus, the study of cusp structures is closely related to the broader problem of understanding the nature of near-threshold hadrons~\cite{Baru:2003qq,Baru:2004xg,Hanhart:2011jz,Hyodo:2013nka, Guo:2017jvc,Sone:2024nfj,Kinugawa:2024crb}.

% isospin symmetry breaking
In many hadron-scattering systems, isospin symmetry gives rise to two or more thresholds that are very close to each other. In such a situation, threshold cusp structures can appear repeatedly within a narrow energy region. Since cusp structures reflect coupled-channel dynamics near the threshold, their behavior is expected to be closely related to the underlying isospin symmetry and its breaking. From this point of view, studying threshold cusp structures in the presence of isospin breaking is useful not only for understanding the line shape itself but also for future analyses of threshold phenomena and near-threshold states. A suitable system for studying this problem is the coupled-channel $\Lambda N$-$\Sigma N$ system. In the charge $Q=+1$ sector, isospin symmetry leads to two nearby thresholds of the $\Sigma^{+}n$ and $\Sigma^{0}p$ channels. This makes the system suitable for discussing the two cusp structures appearing in the $\Lambda p$ elastic cross section.

% YN interaction
Hyperon--nucleon ($YN$) interactions have long been studied as an extension of nuclear forces to the SU(3) flavor sector and play an important role in understanding hypernuclei and neutron-star physics~\cite{Gal:2016boi,Burgio:2021vgk}. Theoretical studies of hyperon--nucleon interactions were initiated with traditional meson-exchange models~\cite{Nagels:1976xq,Holzenkamp:1989tq,Maessen:1989sx} and are now described within chiral effective field theory (EFT), which has been developed up to next-to-next-to-leading order (N$^{2}$LO)~\cite{Haidenbauer:2013oca,Li:2016paq,Haidenbauer:2019boi,Haidenbauer:2023qhf,Haidenbauer:2025zrr}. Experimentally, in addition to old measurements of scattering cross sections, recent progress includes modern $YN$ scattering experiments~\cite{J-PARCE40:2021qxa,CLAS:2021gur,J-PARCE40:2021bgw,J-PARCE40:2022nvq} and studies based on femtoscopy~\cite{ALICE:2018ysd,ALICE:2019buq,ALICE:2021njx,Mihaylov:2023ahn}. Furthermore, the study of the $\Sigma N$ cusp in the in-flight $K^{-}d\to p\Lambda\pi^{-}$ reaction has been proposed as the J-PARC E90 experiment~\cite{Ichikawa:2022pjm}.

% section 2
In this work, we study the isospin-breaking effects on threshold cusp structures. We first introduce a representation of the scattering amplitude in multichannel scattering using the $K$-matrix formalism. This representation enables us to derive transparent expressions for the slopes of the cross section at threshold and thereby to classify the possible cusp structures. We then discuss cusp structures in two-channel and three-channel systems, and show that the possible cusp structures are restricted in these cases. We next examine the threshold cusp behavior of the Flatt\'e amplitude in the same notation as our general amplitude and show that the possible cusp structures generated by the Flatt\'e amplitude are limited. Furthermore, we discuss three-channel scattering with two nearby thresholds, and show that the cusp structures at these thresholds are related when the threshold splitting is small. 

% section 3
Based on this general analysis, we apply the formalism to the $\Lambda p$ elastic cross section in the coupled $\Lambda p$-$\Sigma^+ n$-$\Sigma^0 p$ system. We first show that, when the isospin-breaking effects are small, the relation between the two cusp structures at the $\Sigma N$ thresholds is constrained by isospin symmetry. We then perform quantitative calculations using chiral-EFT input and discuss how isospin breaking modifies the cusp structures.
Preliminary results of this work have been partially reported in conference proceedings~\cite{Sone:2025xuh,Sone:2026qkr}.

%%%%%%%%%%%%%%%%%%%%%%%%%%%%%%%%%%%%%%%%%%%%%%%%
\section{Cusp structures}\label{sec : cusp}
%%%%%%%%%%%%%%%%%%%%%%%%%%%%%%%%%%%%%%%%%%%%%%%%

In this section, we formulate the framework to describe the threshold cusp structures in multi-channel scattering. We first introduce a general $N$-channel scattering amplitude near the threshold of the $N$-th channel using the $K$-matrix representation, and express the near-threshold cross section in terms of the scattering length $a_N$ and the complex constants $b_{ij}$. This allows us to classify the possible cusp structures by the signs of the slopes of the cross section above and below the threshold. We then discuss several specific cases, including two- and three-channel scattering, the Flatt\'e amplitude, and the limit in which two nearby thresholds become degenerate.

%%%%%%%%%%%%%%%%%%%%%%%%%%%%%%%%%%%%%%%%%%%%%%%
\subsection{$N$-channel scattering amplitude}
\label{subsec: N channel}
%%%%%%%%%%%%%%%%%%%%%%%%%%%%%%%%%%%%%%%%%%%%%%%

To discuss the cusp structures, we introduce the $N$-channel $s$-wave two-body scattering amplitude as a function of the energy $E$. Hereafter, we label the scattering channels as $1,2,\ldots,N$ in increasing order of their threshold energies. We focus on the near-threshold energy region of channel $N$ and set the origin of the energy at its threshold. The relative momentum for channel $i$ is denoted by $p_{i}$. 
Near the threshold of channel $N$, $p_{N}(E)=\sqrt{2\mu_{N}E+i0^{+}}$ becomes small and primarily governs the energy dependence, while the momenta of the other channels $(i=1,2,\cdots,N-1)$ can be treated as constants, $p_i=\sqrt{-2\mu_{i}\Delta_{i}}$. Here, $\mu_{i}$ and $\Delta_{i}<0$ denote the reduced mass and the threshold energy of channel $i$, respectively.

Based on the optical theorem, the general form of the $N$-channel scattering amplitude $f^{(N)}(E)$ is given by~\cite{Badalian:1981xj}
\begin{align}
    f^{(N)}(E)
    &=\qty[\hat{K}^{-1} - i\hat{p}(E)]^{-1} 
%    \qty[f^{(N)}(E)]^{-1} &= \hat{K}^{-1} - i\hat{p}, 
    \label{eq: f^N} \\
    &=\hat{K}\qty[1 - i\hat{p}(E)\hat{K}]^{-1} 
    \label{eq: f^N2} ,\\
    \hat{K} &\equiv
    \begin{pmatrix}
        K_{11} & K_{12} & \cdots & K_{1N} \\
        K_{12} & K_{22} & & \\
        \vdots & & \ddots & \vdots\\
        K_{1N} & & \cdots & K_{NN}
    \end{pmatrix}, \label{eq: define the K-mat} 
    \\
    \hat{p}(E)
    &\equiv
    \begin{pmatrix}
        p_{1}                                              \\
         & p_{2}         &        & 0    \\
         &                 & \ddots                     \\
        &0 &        & p_N(E)
\end{pmatrix}, \label{eq: define tha mom-mat}
\end{align}
where $\hat{K}$ is the $N\times N$ real symmetric matrix called the $K$-matrix. 
While the components of the $K$-matrix, $K_{ij}\ (i,j=1,2,\cdots,N)$, generally depend on the energy $E$, in this work, we treat $K_{ij}$ as constants, which corresponds to the leading-order approximation of the momentum expansion near the threshold. In this case, $f^{(N)}(E)$ in Eq.~\eqref{eq: f^N} contains $N(N+1)/2$ independent parameters $K_{ij}$. 
Note that the expression in Eq.~\eqref{eq: f^N} is valid only when $\hat{K}^{-1}$ exists, whereas that in Eq.~\eqref{eq: f^N2} remains applicable even in cases where $\hat{K}^{-1}$ is not defined~\cite{Sone:2024nfj}.

When singularities, such as a CDD zero~\cite{Castillejo:1955ed,Baru:2010ww,Kamiya:2017pcq} or a left-hand cut~\cite{Chew:1960iv, Eden:1966dnq, Du:2024gzw, Wang:2026ups}, lie near the threshold, the leading-order approximation of the $K$-matrix~\eqref{eq: define the K-mat} may become less reliable because the $K$-matrix can exhibit a sizable energy dependence and the radius of convergence of the momentum expansion is correspondingly restricted. 
Nevertheless, if one restricts the discussion to a sufficiently narrow energy region compared with the nearest singularity, the $K$-matrix can still be approximated by a constant at leading order. 
As shown below, the cusp structure is determined by the leading-order term in the channel momentum $p_N(E)$. Therefore, the energy dependence associated with such singularities enters only through higher-order corrections in $p_N(E)$ and does not modify the leading cusp behavior.

From Eq.~\eqref{eq: f^N}, we obtain the $N$-channel scattering amplitude
\begin{align}
    f_{ij}^{(N)}(E) 
    &=
    \frac{n_{ij}(E)}{d(E)} \quad (i,j=1,2,\cdots, N), \label{eq: f by d and n} \\
    d(E) &= \det \qty[\hat{M}(E)], \label{eq: d and M} \\
    n_{ij}(E) &= \tilde{M}_{ij}(E)=(-1)^{i+j} \det \qty[\hat{M}_{\qty{i},\qty{j}}(E)], \label{eq: num rep by M} \\
    \hat{M}(E) &\equiv 
    \hat{K}^{-1} - i\hat{p}(E), \label{eq: def of M}
    %{\rm cof} \qty[f^{N}(E)]^{-1},
    %余因子の書き方を気をつける
\end{align}
where $\tilde{M}(E)$ is the cofactor matrix of $\hat{M}(E)$ and $\hat{M}_{\qty{i},\qty{j}}(E)$ represents the submatrix obtained by deleting the $i$-th row and $j$-th column from $\hat{M}(E)$. We note that the denominator $d(E)$ is common to all components of $f^{(N)}(E)$ whereas the numerator $n_{ij}(E)$ is different for each component. Here, we introduce the constant matrix
\begin{align}
    \hat{R} &\equiv
     \begin{pmatrix}
        [\hat{K}^{-1}]_{11}-ip_1 & [\hat{K}^{-1}]_{12} & \cdots & [\hat{K}^{-1}]_{1N}\\
        [\hat{K}^{-1}]_{12} & [\hat{K}^{-1}]_{22} - ip_2 & \cdots & [\hat{K}^{-1}]_{2N}\\
        \vdots & \vdots & \ddots & \vdots \\
        [\hat{K}^{-1}]_{1N} & [\hat{K}^{-1}]_{2N} & \cdots & [\hat{K}^{-1}]_{NN}
    \end{pmatrix} \label{eq: def of R},
%\\
%    &=
%    \hat{M}(E) 
%    +
%    \begin{pmatrix}
%        0                                              \\
%         & 0        &        & 0    \\
%         &                 & \ddots                     \\
%        &0 &        & ip_N(E)
%\end{pmatrix}
\end{align}
which is related to $\hat{M}(E)$ as
\begin{align}
    \hat{M}(E) 
    &= 
    \hat{R} +
    \begin{pmatrix}
        0                                              \\
         & 0        &        & 0    \\
         &                 & \ddots                     \\
        &0 &        & ip_N(E)
    \end{pmatrix} .
    \label{eq: rel M and R}
\end{align}
Equivalently, $\hat{R}$ corresponds to $\hat{M}(E)$ at the threshold, 
\begin{align}
    \hat{R}=\hat{M}(E=0).
\end{align}
Then, the denominator $d(E)$ can be given by the constant and the linear term in $p_{N}(E)$ as
%The determinant of $\qty[f^{N}(E)]^{-1}$ can be represented by the complex constants~$d_1,d_2$:
\begin{align}
    d(E) &= \det[\hat{R}](1 + ia_Np_N(E)), 
    \label{eq: d by d1 aN}\\
    a_N &\equiv - \det\qty[\hat{M}_{\qty{N},\qty{N}}]/\det[\hat{R}] 
    \label{eq: slength aN M and R}\\
    &=- \det\qty[\hat{R}_{\qty{N},\qty{N}}]/\det[\hat{R}],
\end{align}
where $a_{N}$ represents the scattering length of channel $N$ as shown below. The imaginary part of $a_{N}$ must be negative due to the unitarity. 
We note that $\hat{M}_{\qty{N},\qty{N}}$ does not depend on the energy $E$, because $-ip_N(E)$ in the $N$-th column is deleted from the matrix $\hat{M}(E)$. As a consequence, we obtain $\hat{M}_{\qty{N},\qty{N}}=\hat{R}_{\qty{N},\qty{N}}$.
%Here, we note that $M_{\qty{N},\qty{N}}$ is the constant because it does not depend on $p_N(E)$.  
%This can be checked by expressing $a_N$ in terms of the $K$-matrix components and the relative momenta $p_i\ (i=1,2,\cdots,N-1)$.

First, we discuss the scattering amplitude excluding channel $N$, focusing on the $ij$ components with $i,j=1,\dots,N-1$. 
%In this calculation, we need to consider the diagonal component $d_{ii}(E)$ and the off-diagonal one $d_{ij}(E)\ (i\neq j)$ separately.
Using the matrices $\hat{M}(E)$ and $\hat{R}$, from Eq.~\eqref{eq: num rep by M}, the numerator $n_{ij}(E)\ (i,j=1,2,\cdots,N-1)$ is represented as
\begin{align}
     n_{ij}(E)
%     &= (-1)^{i+j} \det \qty[M_{\qty{i},\qty{j}}(E)] \notag \\
%     &= (-1)^{i+j} \qty{\det\qty[R_{\qty{i},\qty{j}}] - i\det\qty[M_{\qty{iN},\qty{jN}}]p_N(E)} \notag \\
     &= 
     \tilde{R}_{ij}
     \qty(1 + ib_{ij}p_N(E)), 
     \label{eq: num of ij} \\
     b_{ij} &\equiv - \det\qty[\hat{M}_{\qty{iN},\qty{jN}}]/\det\qty[\hat{R}_{\qty{i},\qty{j}}]
     \label{eq: def of bij}\\ 
     &=
     - \det\qty[\hat{R}_{\qty{iN},\qty{jN}}]/\det\qty[\hat{R}_{\qty{i},\qty{j}}], \\
     &\quad (i,j=1,2,\cdots,N-1), \notag
\end{align}
where $\hat{M}_{\qty{iN},\qty{jN}}$ is obtained by deleting the $N$-th row and column from $\hat{M}_{\qty{i},\qty{j}}(E)$. For $N=2$, we define $\hat{M}_{\qty{12},\qty{12}}=1$. In analogy with $\hat{M}_{\qty{N},\qty{N}}$, $\hat{M}_{\qty{iN},\qty{jN}}$ is also independent of energy and equivalent to $\hat{R}_{\qty{iN},\qty{jN}}$. The complex constant $b_{ij}$ determines the energy dependence of the numerator. We note that, from Eqs.~\eqref{eq: slength aN M and R} and~\eqref{eq: def of bij}, the imaginary part of $b_{ii}$ must be negative, because $b_{ii}$ in the $N$-channel case essentially corresponds to the scattering length in the $(N-1)$-channel scattering. On the other hand, the imaginary part of $b_{ij}\ (i\neq j)$ can be positive.  From Eqs.~\eqref{eq: num of ij} and~\eqref{eq: def of bij}, we can represent the $(i,j)$ component of the scattering amplitude using $a_N$ and $b_{ij}$:
\begin{align}
    f^{(N)}_{ij}(E) &= f^{(N)}_{ij}(0) \frac{1+ib_{ij}p_N(E)}{1+ia_Np_N(E)}
    \label{eq: fij by cij aN}\\
    &\quad (i,j=1,2,\cdots,N-1), 
    \notag\\
    f^{(N)}_{ij}(0) &\equiv 
     \tilde{R}_{ij}/\det[\hat{R}]
     =[R^{-1}]_{ij}, 
    \label{eq: def of fij0}
\end{align}
where $f^{(N)}_{ij}(0)$ corresponds to the amplitude at the threshold of channel $N$, namely at $E=0$. For the scattering amplitude considered in this study, the expression in Eq.~\eqref{eq: fij by cij aN} is obtained exactly. For more general scattering amplitudes with an energy-dependent $K$-matrix, $\mathcal{O}(p_N^{2})$ corrections appear in both the numerator and the denominator; however, the leading-order contribution can always be written in the form of Eq.~\eqref{eq: fij by cij aN}.

Next, we study the behavior of the $(i,N)\ (i=1,2,\cdots,N-1)$ component of the scattering amplitude $f^{(N)}_{iN}(E)$ near the threshold of channel $N$. From Eqs.~\eqref{eq: define tha mom-mat} and~\eqref{eq: num rep by M}, the numerator $n_{iN}$ is given by 
\begin{align}
    n_{iN} &= (-1)^{i+N} \det\qty[\hat{R}_{\qty{i},\qty{N}}]=\tilde{R}_{iN} 
    \label{eq: fN iN}\\
    &\quad (i=1,2,\cdots N-1).\notag 
\end{align}
Again, $\hat{R}_{\qty{i},\qty{N}}$ is an energy-independent constant, because the $N$-th column is deleted. Therefore, unlike Eq.~\eqref{eq: fij by cij aN}, $n_{iN}$ does not contain a term proportional to $p_{N}(E)$. From Eq.~\eqref{eq: fN iN}, we obtain the $(i,N)$ component of the scattering amplitude $f^{(N)}_{iN}(E)$ 
\begin{align}
    f^{(N)}_{iN}(E) &= f^{(N)}_{iN}(0)\frac{1}{1+ia_Np_N(E)}. \label{eq: fiN by aN}\\
    &\quad (i= 1,2,\cdots,N-1)
    \notag \\
    f^{(N)}_{iN}(0) &\equiv 
    [R^{-1}]_{iN}, 
    \label{eq: def of fij0}
\end{align}
From Eq.~\eqref{eq: fiN by aN}, it can be seen that the numerator of $f^{(N)}_{iN}(E)$ does not depend on the energy $E$. 

Finally, we consider the behavior of the $(N,N)$ component $f^{(N)}_{NN}(E)$. According to Eq.~\eqref{eq: num rep by M}, the numerator $n_{NN}$ is given by a constant
\begin{align}
    n_{NN} &= \det\qty[\hat{R}_{\qty{N},\qty{N}}]=\tilde{R}_{NN},
\end{align}
and hence $f^{(N)}_{NN}(E)$ can be written as
\begin{align}
    f^{(N)}_{NN}(E) 
    %&= \frac{\det\qty[M_{\qty{N},\qty{N}}]}{\det [R]  - ip_{N}(E) \det\qty[M_{\qty{N},\qty{N}}] } \notag \\
    &= \frac{1}{- \frac{1}{a_N} - ip_N(E)}.  \label{eq: fNNN}
\end{align}
Equation~\eqref{eq: fNNN} shows that the $(N,N)$ component of the scattering amplitude $f^{(N)}_{NN}(E)$ can be written in the form of the effective-range expansion near the threshold of channel $N$~\cite{Sone:2024nfj}, and that the constant $a_{N}$ defined in Eq.~\eqref{eq: slength aN M and R} is identified as the scattering length of channel $N$.
On the other hand, while the off-diagonal components involving channel $N$, $f^{(N)}_{iN}(E)$, exhibit the same energy dependence, their normalization at the threshold differs from that of the $(N,N)$ component. 
%For these components, the numerator contains a linear term in $p_{N}(E)$, and hence the standard effective range expansion is not applicable.

In the discussion so far, we have introduced the scattering length $a_{N}$ and the complex constants $b_{ij}$ in the scattering amplitude $f^{(N)}(E)$. However, $a_{N}$ and $b_{ij}$ are fully determined by the $N(N+1)/2$ independent parameters in the $K$-matrix and are therefore not independent.

%%%%%%%%%%%%%%%%%%%%%%
\subsection{General behavior of cusp structure}
\label{eq: cusp}
%%%%%%%%%%%%%%%%%%%%%%%

In this section, we study the general behavior of cusp structures at the threshold of channel $N$ using the scattering amplitude $f^{(N)}(E)$ introduced in Sec.~\ref{subsec: N channel}. For the components of the amplitude involving channel $N$, the corresponding scattering processes do not occur below the threshold of channel $N$. Therefore, cusp structures cannot be discussed for these components, and we restrict ourselves to the $(i,j)$ components with $i,j=1,2,\cdots,N-1$.

\begin{figure}[tbp]
    \centering
    \includegraphics[width = 9cm, clip]{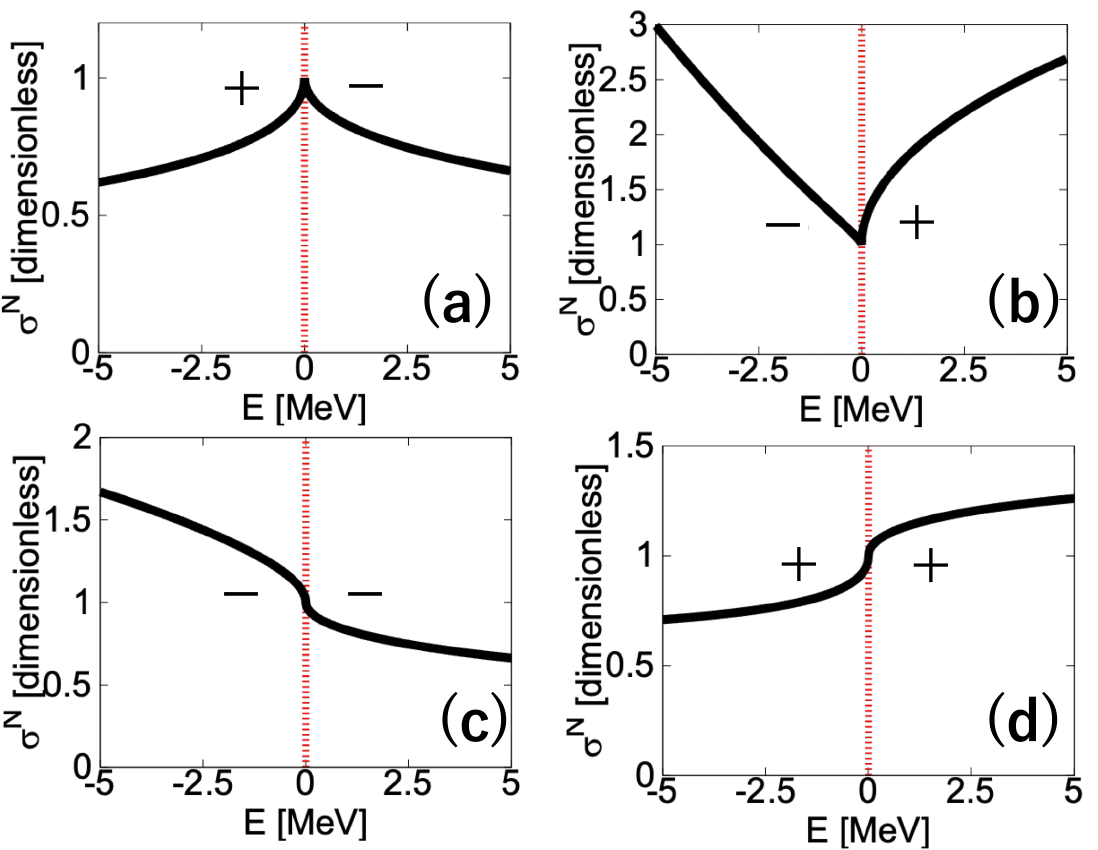}
    \caption{
    %The four kinds
    Four types of the cusp structures in the cross section 
    %$\sigma^{N}(E)$
    $\sigma^{(N)}(E)$.
    The ``$\pm$'' signs above and below the threshold in the figure indicate $\operatorname{sgn}(\Im[a_N-b_{ij}])$ and $-\operatorname{sgn}(\Re[a_N-b_{ij}])$, respectively.
    }
    \label{fig:four kinds}
\end{figure}

In this analysis, we focus on the $s$-wave cross section, which exhibits a cusp at the threshold. To analyze the cusp structure, we define the cross section normalized at the threshold of channel $N$ as
\begin{align}
    \sigma_{ij}^{(N)}(E)
    &\equiv
    \qty|f_{ij}^{(N)}(E)|^2/\qty|f_{ij}^{(N)}(0)|^2
    =\qty|\frac{1+ib_{ij}p_N(E)}{1+ia_Np_N(E)}|^{2}.
    \label{eq: csec swave}
\end{align}
The normalized cross section $\sigma^{(N)}_{ij}(E)$ can be expanded in terms of the momentum of channel $N$ as 
\begin{align}
    \sigma^{(N)}_{ij}(E)
    &=
    1+2\Im[a_N-b_{ij}]p_N(E)+\mathcal{O}(p_N^2(E)),
    \label{eq: slope ij above}\\
    p_N(E)
    &\equiv
    \sqrt{2\mu_NE} \quad (E>0), \\
    \sigma^{(N)}_{ij}(E)
    &=
    1+2\Re[a_N-b_{ij}]\kappa_N(E)+\mathcal{O}(\kappa_N^2(E)),
    \label{eq: slope ij below}\\
    \kappa_N(E)
    &\equiv
    \sqrt{2\mu_N|E|}
    =\sqrt{-2\mu_NE} \quad (E<0). 
\end{align}
Since $p_N(E)$ becomes purely imaginary below the threshold of channel $N$, we use the real quantity $\kappa_N(E)$ in Eq.~\eqref{eq: slope ij below}. When the energy $E$ is taken as the variable, the slopes of the cross section $|d\sigma^{(N)}_{ij}(E)/dE|$ diverge at the threshold. Since the leading nontrivial behavior of the cross section near the threshold is governed by the linear terms in Eqs.~\eqref{eq: slope ij above} and \eqref{eq: slope ij below}, the cusp shape is determined by the signs of their coefficients. Namely, the cusp structures are classified by the signs of $\Im[a_N-b_{ij}]$ and $\Re[a_N-b_{ij}]$. 
Note that the sign of the slope above the threshold is given by
$\operatorname{sgn}(\Im[a_N-b_{ij}])$, whereas that below the threshold is given by
$-\operatorname{sgn}(\Re[a_N-b_{ij}])$. This is because the additional minus sign arises from the derivative of $\kappa_N(E)$ with respect to $E$:
\begin{align}
    \frac{d\sigma^{(N)}_{ij}(E)}{dE}
    &\approx
    2\Re[a_N-b_{ij}]
    \frac{d\kappa_N(E)}{dE}
    ,\notag\\
    &=
    -2\Re[a_N-b_{ij}]\frac{\mu_N}{\kappa_N(E)},
\end{align}
for $E<0$.
The four possible cusp types are shown in Fig.~\ref{fig:four kinds}~\cite{LandauLifshitzQM}.

Although four types of the cusp are possible in the general case, additional constraints arise in two- and three-channel scattering and restrict the allowed cusp structures. We first consider the constraint in two-channel scattering with $N=2$. In two-channel scattering, the cusp structure in the $(1,1)$ component of the cross section is observed at the threshold of channel 2. In this case, according to Eq.~\eqref{eq: def of bij}, we obtain the $(1,1)$ component of the scattering amplitude in the two-channel case
\begin{align}
    f_{11}^{(2)}(E)
    &=
    f_{11}^{(2)}(0) \frac{1+ib_{11}p_2(E)}{1+ia_2p_2(E)}
    , \label{eq: 11 amp} \\
    b_{11} &= - 1/\det\qty[\hat{R}_{\qty{1},\qty{1}}],
\end{align}
where we have used $\det\qty[\hat{M}_{\qty{12},\qty{12}}]=1$. The determinant of the submatrix $\det\qty[\hat{R}_{\qty{1},\qty{1}}]$ can be represented by the $K$-matrix components $K_{ij}$ as
\begin{align}
    \det\qty[\hat{R}_{\qty{1},\qty{1}}]
    &=
    [\hat{K}^{-1}]_{22}
    =
    \frac{K_{11}}{K_{11}K_{22}-K_{12}^2}.
\end{align}
Thus, $b_{11}$ for the two-channel case is given by the real constant
\begin{align}
    b_{11} 
    &=
    -\frac{K_{11}K_{22}-K_{12}^2}{K_{11}}.
    \label{eq: b11 for 2ch}
\end{align}
This is because the imaginary part $-ip_1$, which is the only source of the imaginary part, does not appear in $\hat{R}_{\qty{1},\qty{1}}$. In this way, while the constant $b_{ij}$ is generally complex, the imaginary part of  $b_{11}$ is zero in two-channel scattering. This constraint affects the shape of the cusp structure in $\sigma_{11}^{(2)}(E)$. From Eqs.~\eqref{eq: slope ij above} and~\eqref{eq: b11 for 2ch}, the cross section $\sigma^{(2)}_{11}(E)$ above the threshold is expanded as
\begin{align}
    \sigma^{(2)}_{11}(E)
    &=
    1+2\Im[a_2]p_2(E)+\mathcal{O}(p_2^2(E)), \quad (E>0).
    \label{eq: expansion 2}
\end{align}
From Eq.~\eqref{eq: expansion 2}, we see that the slope of the cross section above the threshold is determined only by the imaginary part of the scattering length $a_2$ because $\Im[b_{11}]$ is absent. In this case, the types of the cusp structures (b) and (d) in Fig.~\ref{fig:four kinds} are forbidden and $\sigma^{(2)}_{11}(E)$ exhibits only the types (a) and (c), because the optical theorem requires $\Im[a_2]<0$. In this way, the cusp structure at the second lowest threshold is restricted to types (a) and (c).

The two-channel scattering amplitude  is also derived within an
effective field theory (EFT) with contact interactions in
Ref.~\cite{Cohen:2004kf}. In their notation, the inverse of the scattering
amplitude is given by
\begin{align}
    [f^{\rm EFT}(E)]^{-1}
    =
    \begin{pmatrix}
        -\dfrac{1}{a_{11}} - ip_1 & \dfrac{1}{a_{12}}\\
        \dfrac{1}{a_{12}} & -\dfrac{1}{a_{22}} - ip_2(E)
    \end{pmatrix}.
    \label{eq: EFT amp}
\end{align}
Comparing Eq.~\eqref{eq: EFT amp} with the constant $K$-matrix representation
in Eq.~\eqref{eq: f^N}, we find that the two amplitudes have the same
form under the identifications
\begin{align}
    [K^{-1}]_{11}=-\frac{1}{a_{11}},\quad
    [K^{-1}]_{22}=-\frac{1}{a_{22}},\quad
    [K^{-1}]_{12}=\frac{1}{a_{12}}.
\end{align}
Thus, the two-channel scattering amplitude used in the present analysis
is consistent with that obtained from the contact EFT in
Ref.~\cite{Cohen:2004kf}. The threshold cusp behavior was also discussed
in Ref.~\cite{Cohen:2004kf}, where the cusp structure of type (c) in Fig.~\ref{fig:four kinds} was identified.

We also show that a similar constraint arises in the inelastic cross section $\sigma^{(3)}_{12}(E)$ in three-channel scattering. From Eq.~\eqref{eq: def of bij}, the scattering amplitude for the $(1,2)$ component in the three-channel scattering is given by
\begin{align}
    f_{12}^{(3)}(E)
    &=
    f^{(3)}_{12}(0)
    \frac{1+ib_{12}p_3(E)}{1+ia_3p_3(E)}, 
    \label{eq: 12 amp}\\
    b_{12}
    &\equiv - \det\qty[\hat{M}_{\qty{13},\qty{23}}]/\det\qty[\hat{R}_{\qty{1},\qty{2}}]
     \label{eq: b12}.
\end{align}
From Eq.~\eqref{eq: def of M}, $\det\qty[\hat{M}_{\qty{13},\qty{23}}]$ is represented by the $K$-matrix components
\begin{align}
    \det\qty[\hat{M}_{\qty{13},\qty{23}}]
    &=
    [\hat{K}^{-1}]_{12}
    =
    \frac{K_{13}K_{23}-K_{12}K_{33}}{\det[\hat{K}]}.
    \label{eq: detM12}
\end{align}
Similarly, $\det\qty[\hat{R}_{\qty{1},\qty{2}}]$ in Eq.~\eqref{eq: b12} can also be represented by the $K$-matrix components $K_{ij}$ 
\begin{align}
    \det\qty[\hat{R}_{\qty{1},\qty{2}}]
    &=
    \frac{K_{12}}{\det[\hat{K}]}.
    \label{eq: detR12}
\end{align}
Then, we obtain the expression of $b_{12}$
\begin{align}
    b_{12}
    &=
    -\frac{K_{13}K_{23}-K_{12}K_{33}}{K_{12}},
    \label{eq: b12 Kmat}
\end{align}
which shows that the constant $b_{12}$ is real, as in the case of $b_{11}$ in Eq.~\eqref{eq: b11 for 2ch} for two-channel scattering. This is because the complex constants $ip_1$ and $ip_2$ do not appear in ${\rm det} [\hat{M}_{\qty{13},\qty{23}}]$ and ${\rm det} [\hat{R}_{\qty{1},\qty{2}}]$. As a result, because $b_{12}$ is real and $\Im[a_3]<0$ from unitarity, the slope of the cross section $\sigma^{(3)}_{12}(E)$ above the threshold of channel 3 is always negative. 
Therefore, $\sigma^{(3)}_{12}(E)$ shows only the cusp structures (a) and (c) in Fig.~\ref{fig:four kinds}.
For the same reason, $\sigma_{21}^{(3)}(E)$ does not exhibit cusp structures of types (b) and (d).

In summary, at a generic threshold, all four types of cusp structures, (a), (b), (c), and (d) in Fig.~\ref{fig:four kinds}, can in principle appear. However,
\begin{itemize}
\item for $\sigma_{11}^{(2)}(E)$ in two-channel scattering (or at the second lowest threshold), and
\item for $\sigma_{12}^{(3)}(E)$ and $\sigma_{21}^{(3)}(E)$ in three-channel scattering (or at the third lowest threshold),
\end{itemize}
only the two types (a) and (c) are present.

%%%%%%%%%%%%%%%%%%%%%%%%%%%
\subsection{Flatt\'{e} amplitude} 
\label{subsec: Flatte}
%%%%%%%%%%%%%%%%%%%%%%%%%%%

In this section, we discuss the cusp structures generated by the Flatt\'e amplitude~\cite{Flatte:1976xu} near the threshold of channel $N$. 
The threshold cusp behavior of the Flatt\'e amplitude has been studied in previous works, for example in Ref.~\cite{Baru:2004xg}. 
Here, we revisit this problem using the notation introduced in Sec.~\ref{subsec: N channel} and clarify the relation between the Flatt\'e amplitude and the $K$-matrix-type scattering amplitude $f^{(N)}(E)$ in Eq.~\eqref{eq: f^N2}.

The Flatt\'{e} amplitude for the $N$-channel case is given by~\cite{Baru:2004xg}
\begin{align}
    f^{\rm F}_{ij}(E)
    &=
    \frac{g_ig_j}{2E_{\rm BW} - 2E - i\sum_{k=1}^{N}g^2_kp_k(E)}
    \label{eq: Flatte for N}\\
    &\quad (i,j=1,2,\cdots,N),
    \nonumber
\end{align}
where $g_i$ denotes the coupling constant and $E_{\rm BW}$ is the bare energy. We note that the numerator of the Flatt\'e amplitude is independent of the energy $E$ for all components. To relate $f^{(N)}(E)$ to $f^{\rm F}(E)$, we focus on the near-threshold ($p_{N}\approx 0$) region by neglecting the $-2E\propto p_N^2$ term in the denominator of Eq.~\eqref{eq: Flatte for N} 
as well as
the energy dependence of the momenta $p_i(E)$ for $i=1,2,\cdots,N-1$ (see, e.g., Appendix A of Ref.~\cite{Sone:2024nfj}). Then, the Flatt\'e amplitude is written as
\begin{align}
     f^{\rm F}_{ij}(E)
     &=
     \frac{\sqrt{r_ir_j}}{\alpha - i\sum_{k=1}^{N-1}r_kp_k - ip_{N}(E)}, 
     \label{eq: Flatte 1st order}\\
     r_i&\equiv g_i^2/g_N^2, 
     \quad 
     \alpha\equiv 2E_{\rm BW}/g_N^2.
     \label{eq: def of alpha r}
\end{align}
The approximate Flatt\'{e} amplitude in Eq.~\eqref{eq: Flatte 1st order} can be obtained as a special case of the $K$-matrix-type amplitude in Eq.~\eqref{eq: f^N2}. 
Indeed, by choosing the $K$-matrix in the separable form
\begin{align}
    K_{ij}
    =
    \frac{g_i g_j}{2E_{\rm BW}},
    \label{eq: K matrix for Flatte}
\end{align}
and by using the infinite series expansion of Eq.~\eqref{eq: f^N2}, we obtain Eq.~\eqref{eq: Flatte 1st order}. The separable $K$-matrix~\eqref{eq: K matrix for Flatte} includes $N$ parameters $g_i/\sqrt{E_{\rm BW}}\ (i=1,2,\cdots,N)$, while the general $K$-matrix contains $N(N+1)/2$ parameters. Namely, the $N(N-1)/2$ constraints are imposed on the Flatt\'{e} amplitude in the general $K$-matrix type amplitude~\eqref{eq: f^N2}~\cite{Sone:2024nfj}.
%Flatteに行けます

From the $(N,N)$ component of the approximate Flatt\'{e} amplitude in Eq.~\eqref{eq: Flatte 1st order}, the scattering length of channel $N$ is determined as
\begin{align}
    a_{N}^{\rm F} &=-\frac{1}{\alpha - i\sum_{k=1}^{N-1}r_kp_k}.
    \label{eq: Flatte scattering length}
\end{align}
Using $a_{N}^{\rm F}$, the $(i,j)$ component of the Flatt\'{e} amplitude can be rewritten as
\begin{align}
    f^{\rm F}_{ij}(E)
     &=
     f_{ij}^{\rm F}(0)\frac{1}{1 + ia_{N}^{\rm F}p_{N}(E)}, 
     \label{eq: Flatte rep aN}\\
     f_{ij}^{\rm F}(0)&\equiv-a_{N}^{\rm F}\sqrt{r_ir_j}
     \label{eq: f0F},\\
     &\quad(i,j=1,2,\cdots,N-1). \notag
\end{align}
By comparing Eq.~\eqref{eq: Flatte rep aN} with the general expression of $f^{(N)}_{ij}(E)$ in Eq.~\eqref{eq: fij by cij aN}, we find that the Flatt\'{e} amplitude does not contain the linear terms in $p_{N}(E)$ in the numerator. In other words, the constants $b_{ij}$ defined in Eq.~\eqref{eq: def of bij} vanish in the Flatt\'{e} amplitude. 
%\sout{Thus, at the level of the near-threshold expression, the Flatt\'{e} amplitude is effectively represented by the form without the $b_{ij}$ terms, while it is generated from the separable K-matrix.} 
%\com{Thus, the Flatt\'{e} amplitude is obtained by imposing the condition $b_{ij}=0$ in the general form of the $K$-matrix amplitude.}
The linear terms in $p_N(E)$ are associated with the background (non-pole) contribution inherent in the coupled-channel scattering amplitude, as discussed in Ref.~\cite{Sone:2024nfj}. Therefore, the Flatt\'{e} amplitude does not contain background contributions. A detailed discussion of the background term in the amplitude components is given in Ref.~\cite{Sone:2024nfj}.

Next, we discuss the cusp behavior of the cross section obtained from the Flatt\'{e} amplitude in Eq.~\eqref{eq: Flatte rep aN}. Using Eqs.~\eqref{eq: csec swave} and~\eqref{eq: Flatte rep aN}, the near-threshold normalized cross section $\sigma^{\rm F}_{ij}(E)$ $(i,j=1,2,\cdots,N-1)$ is expanded as
\begin{align}
    \sigma^{\rm F}_{ij}(E)
    &=
    1+2\Im[a_N^{\rm F}]p_N(E)+\mathcal{O}(p_N^2(E)), \label{eq: Fslope ij above}\\
    &\quad (E>0), \notag
    \\
    \sigma^{\rm F}_{ij}(E)
    &=
    1+2\Re[a_N^{\rm F}]\kappa_N(E)+\mathcal{O}(\kappa_N^2(E)), \label{eq: Fslope ij below}\\
    &\quad (E<0), \notag
    \\
    &\quad (\ i,j=1,2,\cdots,N-1).\notag 
\end{align}
Equations~\eqref{eq: Fslope ij above} and~\eqref{eq: Fslope ij below} show that the slopes of the cross section above and below the threshold are determined only by the scattering length $a_N^{\rm F}$. With the unitarity constraint $\Im[a_N^{\rm F}]<0$, the slope above the threshold is always negative. Using the classification of cusp structures introduced in Sec.~\ref{eq: cusp}, we find that the possible cusp structures of $\sigma^{\rm F}_{ij}(E)$ are restricted to the types (a) and (c) in Fig.~\ref{fig:four kinds}. This restriction is analogous to that found for $\sigma_{11}(E)$ in two-channel scattering and $\sigma_{12}(E)$ in three-channel scattering. However, in the Flatt\'{e} amplitude, this restriction holds for all components of the cross section in arbitrary $N$-channel scattering.
We also note that the condition $b_{ij}=0$ in the Flatt\'e amplitude implies that the cusp structure is determined solely by the scattering length $a_{N}^{\rm F}$. Conversely, this indicates that, in the analysis of cusp structures in general scattering, it is necessary to take into account not only the scattering length $a_{N}^{\rm F}$ but also the coefficients $b_{ij}$ associated with the background contribution.

%%%%%%%%%%%%%%%%%%%%%%%%%%%%%%%%%%%%%%%%%%%%%%%%
\subsection{Cusp structures for nearly degenerate thresholds
% $\Delta_{23}\to 0$ limit
}\label{subsec : close threshold}
%%%%%%%%%%%%%%%%%%%%%%%%%%%%%%%%%%%%%%%%%%%%%%%%

%たまたま閾値が近い場合をかく

%intro
In applications to hadron scattering, multiple thresholds can become adjacent due to isospin symmetry and its breaking. As an example, in the coupled-channel scattering of $\Lambda p$, $\Sigma^+ n$, and $\Sigma^0 p$ discussed in Sec.~\ref{sec : LN-SN scattering}, the threshold energy difference between $\Lambda p$ and $\Sigma^+ n$ is about $75$ MeV, whereas that between the isospin partners $\Sigma^+ n$ and $\Sigma^0 p$ is only about $2$ MeV. 
To discuss cusp structures at such nearly degenerate thresholds, we study a general three-channel scattering problem with two nearby thresholds, separated from the lowest energy channel. To avoid constraints specific to particular systems, here we do not consider the isospin symmetry for the interaction. With this setup, we study the relation between the two cusp structures and their behavior in the limit of vanishing threshold energy difference.

We consider a three-channel system in which the energy difference between
the thresholds of channels 2 and 3 ($\Delta_{23}>0$) is small, while the threshold of
channel 1 is far from these two thresholds. 
We set the origin of the energy at the midpoint between the thresholds
of channels 2 and 3. Namely, the threshold energies of three channels are
$E=\Delta_{1}$, $\Delta_{2}=-\Delta_{23}/2$, and $\Delta_{3}=+\Delta_{23}/2$, and we assume $|\Delta_{1}|\gg \Delta_{23}$. 
In this setup, the
three-channel scattering amplitude is given by
\begin{align}
    %\qty[f^{(3)}(E)]^{-1} &= \hat{K}^{-1} - i\hat{p}, 
    f^{(3)}(E) &= \qty[\hat{K}^{-1} - i\hat{p}(E)]^{-1},
    \label{eq: inv of f^3 close} \\
    \hat{K} \equiv
    \begin{pmatrix}
        K_{11} & K_{12} & K_{13} \\
        K_{12} & K_{22} & K_{23} \\
        K_{13} & K_{23} &  K_{33}
    \end{pmatrix},
    &\quad 
    \hat{p} \equiv
    \begin{pmatrix}
        p_{1}   & 0 & 0                              \\
        0 & p_{2}(E)          & 0    \\
        0 &     0   & p_3(E)
    \end{pmatrix}.
    \label{eq: define tha mom-mat 3ch}
\end{align}
To discuss the cusp structures at the thresholds of channels 2 and 3, we keep the energy dependence of the momenta of
channels 2 and 3 as
%\begin{align}
%     p_2(E)=\sqrt{2\mu_2(E+\Delta_{23}/2)}, 
%     \quad 
%     p_3(E)=\sqrt{2\mu_3(E-\Delta_{23}/2)},
%     \label{eq: close p2 p3}
%\end{align}
\begin{align}
     p_2(E)&=\sqrt{2\mu_2(E+\Delta_{23}/2)+i0^{+}}, \\ 
     p_3(E)&=\sqrt{2\mu_3(E-\Delta_{23}/2)+i0^{+}},
     \label{eq: close p2 p3}
\end{align} 
and treat both $|p_{2}(E)|$ and $|p_{3}(E)|$ as small quantities. 
On the other hand, the momentum of channel 1 is treated as a constant,
$p_{1}=\sqrt{-2\mu_{1}\Delta_{1}}$.

We discuss the cusp structures of the normalized cross section $\sigma^{(3)}_{11}(E)$ at the thresholds of 
channels 2 and 3 and the relation between these two cusp structures using 
the scattering amplitude in Eq.~\eqref{eq: inv of f^3 close}. For this purpose, we first evaluate the quantities that determine the slopes of $\sigma^{(3)}_{11}(E)$ in Eqs.~\eqref{eq: slope ij above} and~\eqref{eq: slope ij below} at each threshold. At the threshold of channel 2, $E=-\Delta_{23}/2$, the cross section behaves as
\begin{align}
    \sigma^{(3)}_{11}(E)
    &=
    1+2\Im[a_2-b^{\rm 2nd}_{11}]p_2(E)+\mathcal{O}(p_2^2(E)),\nonumber \\
    & \quad (E>-\Delta_{23}/2),
    \label{eq: slope ij above 3}\\
    \sigma^{(3)}_{11}(E)
    &=
    1+2\Re[a_2-b_{11}^{\rm 2nd}]\kappa_2(E)+\mathcal{O}(\kappa_2^2(E)), \nonumber \\
    & \quad (E<-\Delta_{23}/2),
    \label{eq: slope ij below 3}\\
    \kappa_2(E)
    &\equiv
    \sqrt{2\mu_2|E+\Delta_{23}/2|} ,
\end{align}
and
the relevant constants are given by
\begin{align}
    a_2 
    &=
     - \frac{\qty(K_{22} - i\Tilde{K}_{33}p_1)
    + \qty(\Tilde{K}_{11} - i\det[\hat{K}]p_1)\kappa_3}
    {\qty(1 - iK_{11}p_1) + \qty(K_{33} - i\Tilde{K}_{22}p_1)\kappa_3},
    \label{eq: a2 in gen}\\
    b_{11}^{2\rm nd} &= - \frac{\Tilde{K}_{33} + \det[\hat{K}]\kappa_3}
    {K_{11} + \Tilde{K}_{22}\kappa_3},
    \label{eq: slope 2 full}\\
    \kappa_3 &\equiv \sqrt{2\mu_3\Delta_{23}},
    \label{eq: def of kappa}
\end{align}
where the superscript ``2nd'' of $b_{11}^{2\rm nd}$ indicates that this quantity 
is evaluated at the threshold of channel 2. 
At this threshold, the momentum of channel 3 is purely imaginary, $p_3=i\sqrt{2\mu_3\Delta_{23}}$, and therefore we introduce the real quantity $\kappa_3$ defined in Eq.~\eqref{eq: def of kappa}. According to Eqs.~\eqref{eq: slope ij above 3} and~\eqref{eq: slope ij below 3}, the slopes of $\sigma^{(3)}_{11}(E)$ at the threshold of channel 2 are determined by the real and imaginary parts of $a_2-b_{11}^{2\rm nd}$. 
Note however that Eq.~\eqref{eq: slope 2 full} shows $\Im b_{11}^{\rm 2nd}=0$, in accordance with the discussion in Sec.~\ref{eq: cusp} for the second lowest threshold.

Because $\kappa_{3}$ is small when $\Delta_{23}\to 0$, we expand
 $a_2-b_{11}^{2\rm nd}$ in terms of $\kappa_3$ as
\begin{align}
    a_2 - b_{11}^{2\rm nd}
    &= W^{(0)} + W(\kappa_3),
    \label{eq: expansion of slope 2}\\
   W^{(0)} &\equiv - \frac{K_{12}^2}{(1 - iK_{11}p_1)K_{11}}, 
   \label{eq: def of L} \\
    W(\kappa_3)
    &\equiv
    \sum_{n=1}^{\infty}
    \frac{\kappa_3^n}{n!} 
    \left.\frac{d^n(a_2 - b_{11}^{2\rm nd})}{d\kappa_3^n}\right|_{\kappa_3=0}.
    \label{eq: def of W}
\end{align}
Here, $W^{(0)}$ is the leading contribution that remains in the 
$\Delta_{23}\to0$ limit, while $W(\kappa_3)$ represents the correction 
arising from the finite separation between the two thresholds.

We also consider the slopes of the cross section at the threshold of channel 3, 
$E=\Delta_{23}/2$. 
At this threshold, the relevant constants are given by
\begin{align}
    a_3 
    &=
    - \frac{\qty(K_{33} - i\Tilde{K}_{22}p_1) - i\qty(\Tilde{K}_{11} - i\det[\hat{K}]p_1)p_2}
    {\qty(1 - iK_{11}p_1)
    -i\qty(K_{22} - i\Tilde{K}_{33}p_1)p_2},
    \label{eq: a3 in gen}\\
    b_{11}^{\rm 3rd}&= -
    \frac{\Tilde{K}_{22} - i\det[\hat{K}]p_2}{K_{11} - i\Tilde{K}_{33}p_2},
    \label{eq: slope 3 full}\\
    p_2 &= \sqrt{2\mu_2\Delta_{23}}, \label{eq: p2 at third threshold}
\end{align}
where the superscript ``3rd'' of $b_{11}^{\rm 3rd}$ indicates that this quantity 
is evaluated at the threshold of channel 3. 
At this threshold, the momentum of channel 2 is real and is denoted by $p_2$. 
According to Eqs.~\eqref{eq: slope ij above} and~\eqref{eq: slope ij below}, 
the slopes of $\sigma^{(3)}_{11}(E)$ at the threshold of channel 3 are 
determined by the real and imaginary parts of $a_3-b_{11}^{\rm 3rd}$. 
Expanding $a_3-b_{11}^{\rm 3rd}$ in terms of $p_2$, we obtain
\begin{align}
    a_3 - b_{11}^{\rm 3rd}
    &=
    X^{(0)} + X(p_2), 
    \label{eq: expansion of slope 3}\\
    X^{(0)}
    &\equiv - \frac{K_{13}^2}{(1 - iK_{11}p_1)K_{11}}, 
    \label{eq: def of R}\\
    X(p_2)
    &\equiv
    \sum_{n=1}^{\infty}
    \frac{p_2^n}{n!} 
    \left.\frac{d^n(a_3 - b_{11}^{\rm 3rd})}{dp_2^n}\right|_{p_2=0}.
    \label{eq: def of X}
\end{align}
Here, $X^{(0)}$ is the leading contribution and $X(p_2)$ represents the correction 
arising from $\Delta_{23}$.

We now discuss how the cusp structures at the thresholds of channels 2 
and 3 are related when these two thresholds are close to each other. 
If the correction terms $W(\kappa_3)$ and $X(p_2)$ are small compared with 
the leading terms $W^{(0)}$ and $X^{(0)}$, respectively, the slopes of the 
cross section at the two thresholds are determined by $W^{(0)}$ and $X^{(0)}$. 
This situation is realized when $\Delta_{23}$ is sufficiently small, 
because both $\kappa_3$ and $p_2$ vanish in the limit $\Delta_{23}\to0$. 
From Eqs.~\eqref{eq: def of L} and~\eqref{eq: def of R}, the leading terms 
$W^{(0)}$ and $X^{(0)}$ have the same denominator, while their numerators 
differ only by the real positive factors $K_{12}^2$ and $K_{13}^2$. 
Therefore, the signs of the real and imaginary parts of $W^{(0)}$ and 
$X^{(0)}$ are identical:
\begin{align}
    \operatorname{sgn}(\Re[W_{0}])
    &= \operatorname{sgn}(\Re[X_{0}]), \\
    \operatorname{sgn}(\Im[W_{0}])
    &= \operatorname{sgn}(\Im[X_{0}])
\end{align}
As a result, the cross section $\sigma^{(3)}_{11}(E)$ shows the same 
types of cusp structures at the thresholds of channels 2 and 3, provided 
that
\begin{align}
    \left|\frac{W(\kappa_3)}{W^{(0)}}\right| \ll 1,
    \quad
    \left|\frac{X(p_2)}{X^{(0)}}\right| \ll 1.
\end{align}
In this situation, the cusp structure at the threshold of channel 2 is 
restricted to the types (a) and (c) in Fig.~\ref{fig:four kinds}, because the slope above the threshold 
is constrained by the unitarity condition $\Im[a_2]<0$ and $\Im b_{11}^{\rm 2nd}=0$. 
Therefore, the same types of cusp structures, namely the types (a) and 
(c), are expected to appear also at the threshold of channel 3. 
When the correction terms are not negligible, however, this expectation 
does not necessarily hold.
If there exist multiple channel thresholds below the degenerate threshold, all four types of cusps can in principle appear, since no constraint is imposed on the cusp of the lower channel. However, the fact that the same type of cusp appears at the two thresholds is common to the three-channel case.

Next, we consider the behavior of the cusp structures in the limit $\Delta_{23}\to0$.
For finite $\Delta_{23}$, the cross section $\sigma^{(3)}_{11}(E)$ exhibits two cusp structures at the thresholds of channels 2 and 3. In the limit $\Delta_{23}\to0$, these two threshold energies become identical, and only a single cusp appears at the degenerate threshold. We denote the corresponding degenerate channel by channel \ff. The reduced masses of channels 2 and 3 also become identical, $\mu_2=\mu_3\equiv\mu_{\rm \ff}$, and the matrix $\hat{p}$ in Eq.~\eqref{eq: define tha mom-mat 3ch} is reduced to
\begin{align}
    \hat{p} \equiv
    \begin{pmatrix}
        p_{1}   & 0 & 0                              \\
        0 & p_{\rm \ff}(E)          & 0    \\
        0 &     0   & p_{\rm \ff}(E)
    \end{pmatrix}
    \quad (\Delta_{23}\to0),
    \label{eq: pmat del0}
\end{align}
where $p_{\rm \ff}(E)=\sqrt{2\mu_{\rm \ff}E+i0^{+}}$ is the momentum of 
channel \ff. 
Using the $K$-matrix in Eq.~\eqref{eq: define tha mom-mat 3ch} and 
Eq.~\eqref{eq: pmat del0}, the $(1,1)$ component of the scattering 
amplitude in the $\Delta_{23}\to0$ limit is obtained as
\begin{align}
    f^{(3)}_{11}(E)
    &=
    f^{(3)}_{11}(0)\frac{1+iBp_{\rm \ff}(E)}
    {1+iAp_{\rm \ff}(E)}
    \quad (\Delta_{23}\to 0), 
    \label{eq: f(3) in DEL0}
    \\
    f^{(3)}_{11}(0) &\equiv \left.[R^{-1}]_{11}\right|_{\Delta_{23}\to 0},
    \label{eq: norm 3ch}
    \\
    A &\equiv - \frac{K_{22}+K_{33} - i(\Tilde{K}_{22}+\Tilde{K}_{33})p_1}
    {1 - iK_{11}p_1}, 
    \label{eq: def of A}\\
    B &\equiv - \frac{\Tilde{K}_{33}+\Tilde{K}_{22}}{K_{11}}.
    \label{eq: def of B}
\end{align}
In contrast to Eq.~\eqref{eq: fij by cij aN},
 the terms of order $p_{\rm \ff}^2$ arise in both numerator and denominator, which are neglected in 
Eq.~\eqref{eq: f(3) in DEL0}. 
Related to this, the complex constant $A$ should not be identified with the scattering length of channel \ff, because no component of the amplitude in the $\Delta_{23}\to0$ limit can be written in the standard form of the effective range expansion.
The constant $B$ is real, as is the constant $b_{11}$ in two-channel scattering discussed around Eq.~\eqref{eq: b11 for 2ch}.

We next show that the constants $A$ and $B$ in the $\Delta_{23}\to0$ limit 
are obtained from the sums of the corresponding quantities at the thresholds 
of channels 2 and 3. 
Taking the limit $\Delta_{23}\to0$ $(p_{2},\kappa_{3}\to 0)$ in 
Eqs.~\eqref{eq: a2 in gen}, \eqref{eq: slope 2 full}, 
\eqref{eq: a3 in gen}, and~\eqref{eq: slope 3 full}, and comparing the 
result with Eqs.~\eqref{eq: def of A} and~\eqref{eq: def of B}, we find
\begin{align}
    A &= a_2 + a_3 
    \quad (\Delta_{23}\to 0), 
    \label{eq: A sum} \\
    B &= b_{11}^{2\rm nd} + b_{11}^{3\rm rd} 
    \quad (\Delta_{23}\to 0).
    \label{eq: sum B}
\end{align}
In the $\Delta_{23}\to0$ limit, the cusp behavior of 
$\sigma^{(3)}_{11}(E)$ at the threshold of channel \ff is determined by 
the combination $A-B$. 
Equations~\eqref{eq: A sum} and~\eqref{eq: sum B} imply that this 
combination is obtained from the sum of the corresponding leading 
contributions at the thresholds of channels 2 and 3 for finite 
$\Delta_{23}$. 
Using Eqs.~\eqref{eq: A sum} and~\eqref{eq: sum B}, we obtain
\begin{align}
    A - B &= W^{(0)} + X^{(0)} \notag \\
    &= 
    - \frac{K_{12}^2 + K_{13}^2}{(1 - iK_{11}p_1)K_{11}}.
    \label{eq: A - B}
\end{align}
Equation~\eqref{eq: A - B} shows that, in the $\Delta_{23}\to0$ limit, the slopes of the cross section associated with the two separate thresholds are combined into a single slope at the degenerate threshold. In addition, since the numerator $K_{12}^2+K_{13}^2$ is positive, the signs of the real and imaginary parts of $A-B$ are determined only by the common denominator $(1-iK_{11}p_1)K_{11}$. Thus, the cross section $\sigma^{(3)}_{11}(E)$ in the $\Delta_{23}\to0$ limit shows a cusp structure of the same type as the cusp structures at the thresholds of channels 2 and 3 for finite $\Delta_{23}$, when the correction terms $W(\kappa_3)$ and $X(p_2)$ are negligible. $\Im[b_{11}^{\rm 3rd}]$ vanishes due to the degenerate thresholds, and therefore $\Im[B]$ also vanishes. In this case, the types of the cusp structures at the threshold of channel \ff are restricted to (a) and (c), because the slope of the cross section above the threshold is determined only by $\Im[A]<0$ [see Eq.~\eqref{eq: A sum}].

%\com{これも今の場合は(a),(c)のみではないでしょうか。$b_{11}^{\rm 3rd}$の虚部は$p_{3}\to 0$で消えるので、$A-B$の虚部は$A$の虚部になります。effectiveに2チャンネルになっていると言っても良いかもしれません。}

%%%%%%%%%%%%%%%%%%%%%%%%%%%%%%%%%%%%%%%%%%%%%%%%%%%%%%%%
\section{Application to $\Lambda N$-$\Sigma N$ scattering}\label{sec : LN-SN scattering}
%%%%%%%%%%%%%%%%%%%%%%%%%%%%%%%%%%%%%%%%%%%%%%%%%%%%%%%%

In this section, we apply the general discussion in Sec.~II to the $\Lambda N$-$\Sigma N$ system with charge $Q=+1$, namely the coupled $\Lambda p$-$\Sigma^+ n$-$\Sigma^0 p$ system. In this system, the $\Sigma^+ n$ and $\Sigma^0 p$ thresholds are separated only by a small isospin-breaking effect, while both channels are coupled to $\Lambda p$. This makes the system suitable for studying how isospin breaking affects the threshold cusp structures. We first discuss the general constraints imposed by isospin symmetry on the cusp structures in the $\Lambda p$ elastic cross section. We then perform numerical calculations for simplified examples and for realistic input based on chiral effective field theory, and examine how the cusp structures are modified by isospin breaking effects.

%%%%%%%%%%%%%%%%%%%%%%%%%%%%%%%%%%%%%%%%%%%%%%%%%%%%%%%%
\subsection{Cusp structure in the $\Lambda p$ cross section}
\label{subsec: cusp lambda p}
%%%%%%%%%%%%%%%%%%%%%%%%%%%%%%%%%%%%%%%%%%%%%%%%%%%%%%%%

We consider the three scattering channels $\Lambda p$, $\Sigma^+ n$, and 
$\Sigma^0 p$, which are labeled as channels 1, 2, and 3, respectively. 
In this case, since no Coulomb interaction is present in any channel, the scattering amplitude in Sec.~\ref{sec : cusp} can be applied.
The threshold energies of channels 2 and 3 are very close to each other 
because of isospin symmetry, while their small difference arises from 
isospin breaking. 
As in Sec.~\ref{subsec : close threshold}, we denote the energy difference between the thresholds of channels 2 and 
3 by $\Delta_{23}$ and the origin of the energy is chosen at the midpoint between the $\Sigma^{+}n$ and $\Sigma^{0}p$ thresholds. 
With finite $\Delta_{23}$, the isospin broken system is realized, 
whereas the isospin symmetric system is realized in the limit $\Delta_{23}\to0$. 
In the isospin-symmetric limit, we denote the degenerate $\Sigma N$ 
channel by channel \ff.

We study the cusp structures of the $\Lambda p$ elastic cross section, 
namely the $(1,1)$ component, at the thresholds of channels 2 and 3. 
In our analysis, the $K$-matrix for $\Lambda N$-$\Sigma N$ scattering is 
assumed to respect isospin symmetry, which reduces the number of independent 
parameters from six in the general three-channel case~\eqref{eq: define tha mom-mat 3ch} to four. Specifically, we express the $K$-matrix by four real constants $C_{i}$ ($i=1,...,4$) as
\begin{align}
    K
    &=
    \begin{pmatrix}
        C_1 & \sqrt{2}C_4 & C_4 \\
        \sqrt{2}C_4 & C_2 & \sqrt{2}(C_2 - C_3) \\
        C_4 & \sqrt{2}(C_2 - C_3) & C_3
    \end{pmatrix}.
    \label{eq: Kmatrix for lambdaN}
\end{align}
The effect of isospin breaking is therefore incorporated through the threshold energy difference in the momentum matrix $\hat{p}$ rather than through the $K$-matrix. In the present case, the momentum matrix is given by Eq.~\eqref{eq: define tha mom-mat 3ch}, and the threshold separation $\Delta_{23}$ enters the momentum functions $p_2(E)$ and $p_3(E)$ as in Eq.~\eqref{eq: close p2 p3}. In the isospin-symmetric limit $\Delta_{23}\to0$, these momenta coincide with each other, $p_2(E)=p_3(E)=p_{\rm \ff}(E)$, and the momentum matrix reduces to Eq.~\eqref{eq: pmat del0}. Thus, in this limit, the scattering amplitude becomes fully consistent with 
isospin symmetry.
%\com{このときブロック対角にできる話をしておいても良い}

%LHCの話

As discussed in Sec.~\ref{subsec: N channel}, near-threshold singularities restrict the 
convergence region of the momentum expansion. 
In the $\Lambda N$-$\Sigma N$ system, the left-hand cut arising from 
$t$-channel one-pion exchange in $\Sigma N$ scattering lies close to 
the $\Sigma N$ threshold, with the corresponding branch point located 
approximately $5.1$ MeV below the $\Sigma N$ threshold~\cite{Haidenbauer:2021smk}. 
Nevertheless, sufficiently close to the $\Sigma N$ threshold, the
effects of the left-hand cut can be absorbed into the low-energy
parameters, and the cusp classification and the near-threshold analysis
presented in this work remain applicable.
On the other hand, as the energy approaches the left-hand cut branch 
point, the explicit contribution from the cut becomes important and 
the constant-$K$ approximation would break down~\cite{Du:2024gzw, Wang:2026ups}.

First, we consider the isospin-broken case with finite $\Delta_{23}$. 
According to the discussion in Sec.~\ref{subsec : close threshold}, the cusp shapes at the 
thresholds of channels 2 and 3 are determined by the signs of the real 
and imaginary parts of $a_2-b^{2\rm nd}_{11}$ and $a_3-b^{3\rm rd}_{11}$, 
respectively. 
Substituting the isospin-symmetric $K$-matrix in Eq.~\eqref{eq: Kmatrix for lambdaN} 
into the general expressions in Eqs.~\eqref{eq: expansion of slope 2} and~\eqref{eq: expansion of slope 3}, these quantities are expanded as
\begin{align}
    a_2 - b^{2\rm nd}_{11} 
    &= 2R + W(\kappa_{3})
    \quad (\text{2nd threshold}),  
    \label{eq: slope of th 2} \\
    a_3 - b^{3\rm rd}_{11} 
    &= R + X(p_{2})
    \quad (\text{3rd threshold}), 
    \label{eq: slope of th 3} \\
    R &\equiv -\frac{C_4^2}{(1 - i C_1 p_1) C_1}.
    \label{eq: def of R lambda}
\end{align}
Here, the terms $W(\kappa_3)$ and $X(p_2)$ represent 
corrections due to the finite threshold separation. In the limit 
$\Delta_{23}\to0$, $W(\kappa_{3})$ and $X(p_{2})$ vanish, while the leading terms remain finite. 
In the general three-channel case discussed in Sec.~\ref{subsec : close threshold}, 
the two leading contributions are proportional to $K_{12}^2$ and $K_{13}^2$, 
and therefore the relative size of the two cusp coefficients depends on 
the details of the $K$-matrix. 
In the present $\Lambda N$-$\Sigma N$ case, however, isospin symmetry fixes 
the ratio of these couplings as $K_{12}^2/K_{13}^2=2$. 
As a result, the leading contributions in Eqs.~\eqref{eq: slope of th 2} 
and~\eqref{eq: slope of th 3} are given by $2R$ and $R$, respectively. 
Thus, when the correction terms are negligible, the cusp structures at the 
$\Sigma^+ n$ and $\Sigma^0 p$ thresholds are not only of the same type, 
but their relative size in the momentum expansion is also fixed by isospin 
symmetry. 
Although the derivative of the cross section with respect to the energy 
diverges at the threshold, the larger leading coefficient implies that the 
cusp at the $\Sigma^+ n$ threshold changes more rapidly as a function of 
energy than that at the $\Sigma^0 p$ threshold. 
When the correction terms are not negligible, however, this simple relation 
does not necessarily hold.

Finally, we briefly discuss the isospin-symmetric limit, 
$\Delta_{23}\to0$. 
As discussed in Sec.~\ref{subsec : close threshold}, when the two thresholds 
become degenerate, the two cusp structures are combined into a single cusp 
at the threshold of channel \ff. 
For the isospin-symmetric $K$-matrix in Eq.~\eqref{eq: Kmatrix for lambdaN}, 
the quantity that determines the cusp structure in this limit is given by
\begin{align}
    A - B =a_2 +a_3 - b^{2\rm nd}_{11} - b^{3\rm rd}_{11}  = 3R.
    \label{eq: slopes of isospin limit}
\end{align}
This result corresponds to the sum of the leading contributions $2R$ and 
$R$ at the thresholds of channels 2 and 3. 
Thus, the factor 3 in the isospin-symmetric limit reflects the merging of 
the two cusp structures discussed above.

Here, we consider the scattering lengths in the isospin basis. In the $\Lambda N$-$\Sigma N$ system, there are two isospin channels, $I=1/2$ and $I=3/2$. In the $\Delta_{23}\to 0$ limit, the $\Sigma N$ channel couples to $\Lambda N$ in the $I=1/2$ channel, whereas it does not couple to any lower channel in the $I=3/2$ channel. The relations between the scattering lengths in the charge basis and those in the isospin basis are given by
\begin{align}
     a_{1/2} &= - a_3 + 2a_2,
     \label{eq: a2/2 a2 a3}\\
     a_{3/2} &= 2a_3 - a_2,
     \label{eq: a3/2 a2 a3}
\end{align}
where $a_{1/2}$ and $a_{3/2}$ denote the $\Sigma N$ scattering lengths in the $I=1/2$ and $I=3/2$ channels, respectively. 
From Eqs.~\eqref{eq: A sum} and~\eqref{eq: a2/2 a2 a3}, it can be seen that 
the parameter $A=a_{2}+a_{3}$ is different from the scattering length $a_{1/2}$ in the isospin basis. In fact, as we mentioned in Sec.~\ref{subsec : close threshold}, $A$ cannot be identified as the scattering length. Instead, if we take the $I=1/2$ combination of the slope coefficients in Eqs.~\eqref{eq: slope of th 2} and \eqref{eq: slope of th 3} in the $\Delta_{23}\to 0$ limit, we obtain
\begin{align}
     - (a_3-b_{11}^{\rm 3rd}) + 2(a_2-b_{11}^{\rm 2nd}) 
     = 
     3R = A-B
\end{align}
which coincides with the coefficient in Eq.~\eqref{eq: slopes of isospin limit}. Namely, it is the slope coefficient of the $\Lambda p$ elastic cross section that follows the isospin relation, rather than the scattering lengths of corresponding channels.

%\com{From Eqs.~\eqref{eq: A sum} and~\eqref{eq: a2/2 a2 a3}, one can see that $A$ is not identical to $a_{1/2}$ in isospin basis, although the $\Lambda p$ channel couples only to the $\Sigma N$ channel with $I=1/2$. The slopes of the cross section in isospin-symmetric limit are given by
%\begin{align}
%     a_{1/2} - b_{11}^{I=1/2}
%     &= 
%     - (a_3-b_{11}^{\rm 3rd}) + 2(a_2-b_{11}^{\rm 2nd}) \notag \\
%     &= 
%     3R
%\end{align}
%Namely, the slopes of the $\Lambda p$ elastic cross section in the charge basis are identical to that with $I=1/2$ in the isospin basis.}

%%%%%%%%%%%%%%%%%%%%%%%%%%%%%%%%%%%%%%%%%%%%%%%%%%%%%%%%
\subsection{
%Numerical analysis: Flatt\'{e} limit
Numerical results with Flatt\'{e} amplitude
}\label{sec: cs fg}
%%%%%%%%%%%%%%%%%%%%%%%%%%%%%%%%%%%%%%%%%%%%%%%%%%%%%%%%

In this section, we numerically study the threshold cusp structures in $\Lambda N$-$\Sigma N$ scattering. We first consider the simplified case where the scattering amplitude is given by the Flatt\'e amplitude introduced in Sec.~\ref{subsec: Flatte}, namely, the case without a background contribution.  The hadron masses used in the calculation are taken from Ref.~\cite{ParticleDataGroup:2024cfk}.

We use the isospin-symmetric $K$-matrix in Eq.~\eqref{eq: Kmatrix for lambdaN} for the numerical calculation. To realize the Flatt\'e amplitude, we impose the conditions 
\begin{align}
    C_1C_3-C_4^2=0,  \label{eq: condition F1}\\
    C_2-2C_3=0. \label{eq: condition F2}
\end{align}
Under these conditions, the $K$-matrix in Eq.~\eqref{eq: Kmatrix for lambdaN} becomes separable, and the number of independent parameters characterizing the scattering amplitude is reduced from four to two. Equations~\eqref{eq: condition F1} and~\eqref{eq: condition F2} 
provide one of the 
possible separable solutions. Other separable solutions and their interpretation in the isospin basis are discussed in Appendix~\ref{app:separable_K_isospin}.

Under the conditions in Eqs.~\eqref{eq: condition F1} and~\eqref{eq: condition F2}, the scattering lengths of channels 2 and 3 with $\Delta_{23}\to 0$ in Eqs.~\eqref{eq: a2 in gen} and~\eqref{eq: a3 in gen} are given only by $C_1$ and $C_2$:
\begin{align}
    a_2
    =
    2a_3
    =
    - \frac{C_2}
    {1 - iC_1p_1}.
    \label{eq: aF Sigma N}
\end{align}
In this case, $a_2$ and $a_3$ are no longer independent. Substituting Eq.~\eqref{eq: aF Sigma N} into Eqs.~\eqref{eq: a2/2 a2 a3} and~\eqref{eq: a3/2 a2 a3}, we obtain
\begin{align}
    a_{1/2}
    &=
    \frac{3}{2}a_2, \\
    a_{3/2}
    &= 0.
\end{align}
From Eqs.~\eqref{eq: slopes of isospin limit} and \eqref{eq: aF Sigma N}, we obtain $A=a_2+a_3=a_{1/2}$. Under the conditions Eqs.~\eqref{eq: condition F1} and~\eqref{eq: condition F2}, we find that $b_{11}^{\rm 2nd }$ and $b_{11}^{\rm 3rd}$ vanish. Thus, when the $K$-matrix is chosen to be separable so that the scattering amplitude reduces to the Flatt\'e form, cusp structures in the $\Lambda p$ cross section are determined only by $a_{1/2}$.

For the numerical calculation, we choose the $I=1/2$ scattering length as
\begin{align}
    a_{1/2} = -1.0 - i0.8~{\rm fm}, \quad (\Delta_{23}\to0),
    \label{eq: a2s num}
\end{align}
For this value of $a_{1/2}$, a quasivirtual pole~\cite{Nishibuchi:2023acl} is generated below the $\Sigma N$ threshold in the isospin-symmetric $I=1/2$ amplitude. In this simplified model, the four parameters $C_1$, $C_2$, $C_3$, and $C_4$ are uniquely fixed\footnote{As discussed in the Appendix, the sign of $C_{4}$ is not fixed; however, since it does not affect the result, we adopt the solution with $C_{4}>0$ here.} by the value of $a_{1/2}$ together with the two conditions in Eqs.~\eqref{eq: condition F1} and~\eqref{eq: condition F2}. The resulting $K$-matrix parameters are
\begin{align}
    C_1 &= 0.56~{\rm fm},
    \quad
    C_2 = 1.1~{\rm fm}, \\
    C_3 &= 0.55~{\rm fm}, 
    \quad
    C_4 = 0.55~{\rm fm}.
\end{align}
Using these parameters, the scattering lengths in the isospin-broken case are obtained as
\begin{align}
    a_2 &= -0.65-i0.46~{\rm fm}, \quad (\Delta_{23}\neq0),\\
    a_3 &= -0.26-i0.27~{\rm fm}, \quad (\Delta_{23}\neq0).
    \label{eq: cp a3}
\end{align}
The isospin relation $a_2=2a_3$ is slightly broken due to the finite $\Delta_{23}$. Since the present scattering amplitude is of the Flatt\'e form, the cusp structures at the thresholds of channels 2 and 3 are determined solely by the corresponding scattering lengths $a_2$ and $a_3$, as discussed in Sec.~\ref{subsec: Flatte}.

\begin{figure}[tbp]
    \centering
    \includegraphics[width = 9cm, clip]{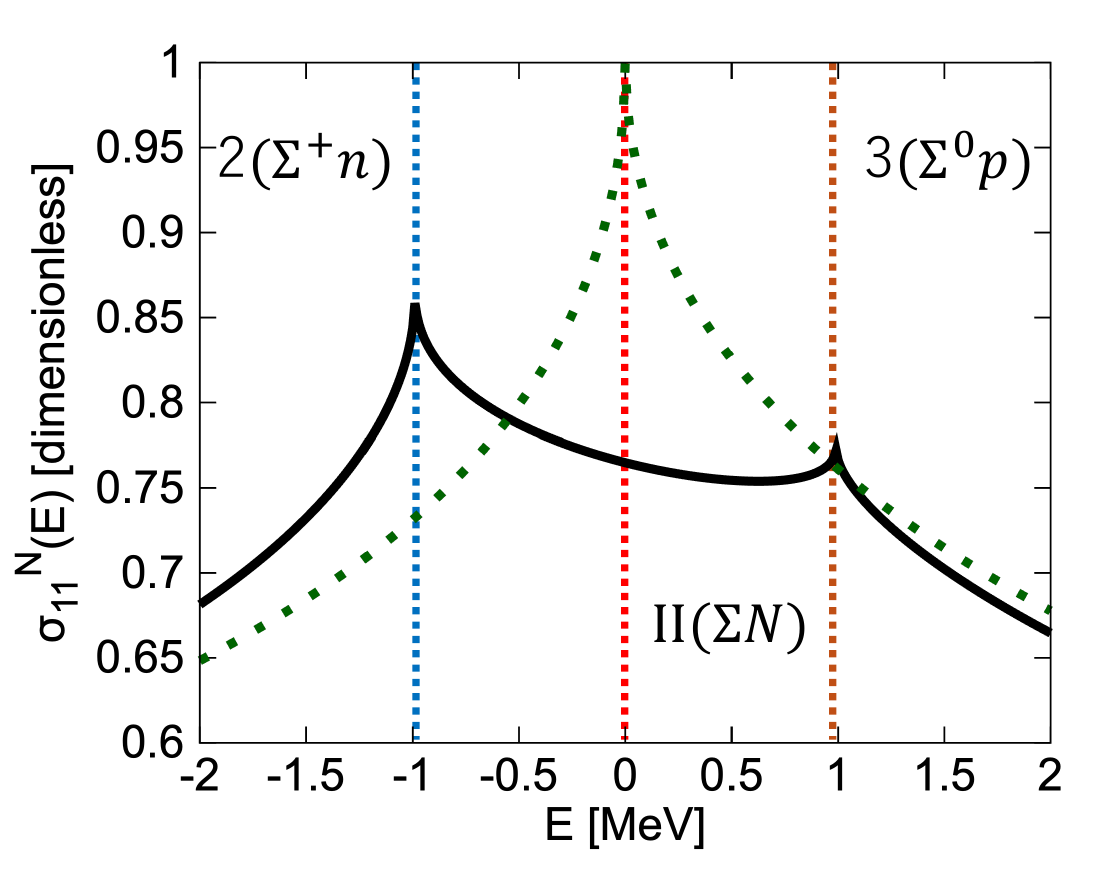}
    \caption{
    The normalized $\Lambda p$ elastic cross section from the Flatt\'{e} amplitude with fixed $a_{1/2}=-1.0 - i0.8~{\rm fm}$. The dotted and solid lines represent the cases with $\Delta_{23}\to0$ and $\Delta_{23}\neq0$, respectively. The vertical dotted lines indicate the thresholds of channels 2, II, and 3.
    }
    \label{fig: F1}
\end{figure}

The normalized cross section corresponding to the parameter choice in Eq.~\eqref{eq: a2s num} is shown in Fig.~\ref{fig: F1}. The cross section with $\Delta_{23}\neq0$ is also normalized by $|f^{(3)}_{11}(0)|^2$ in Eq.~\eqref{eq: norm 3ch}. The dotted and solid lines represent the isospin-symmetric and isospin-broken cases, respectively. The vertical lines represent the thresholds of channels 2, II, and 3. As discussed in Sec.~\ref{subsec : close threshold}, even when two thresholds happen to be close to each other, the cusp structures at the two thresholds are of the same type. In the present $\Lambda N$-$\Sigma N$ system, the solid line indeed shows that the cusp structures at the $\Sigma^{+}n$ and $\Sigma^{0}p$ thresholds are both upward and hence of the same type. In terms of the classification in Fig.~1, these cusp structures correspond to type (a).

Furthermore, the cusp at the $\Sigma^{+}n$ threshold is more pronounced than that at the $\Sigma^{0}p$ threshold. This behavior is consistent with the discussion in Sec.~\ref{subsec: cusp lambda p}, where isospin symmetry relates the leading contributions to the cusp structures at the $\Sigma^{+}n$ and $\Sigma^{0}p$ thresholds as $2R$ and $R$, respectively. These results indicate that the isospin-breaking effects are small in the present example. Indeed, the scattering lengths in the isospin-broken case approximately satisfy $a_2 \approx 2a_3$. Therefore, the cusp shape is mainly determined by the leading term $R$, while the subleading corrections associated with the threshold splitting remain small.

\begin{figure}[tbp]
    \centering
    \includegraphics[width = 9cm, clip]{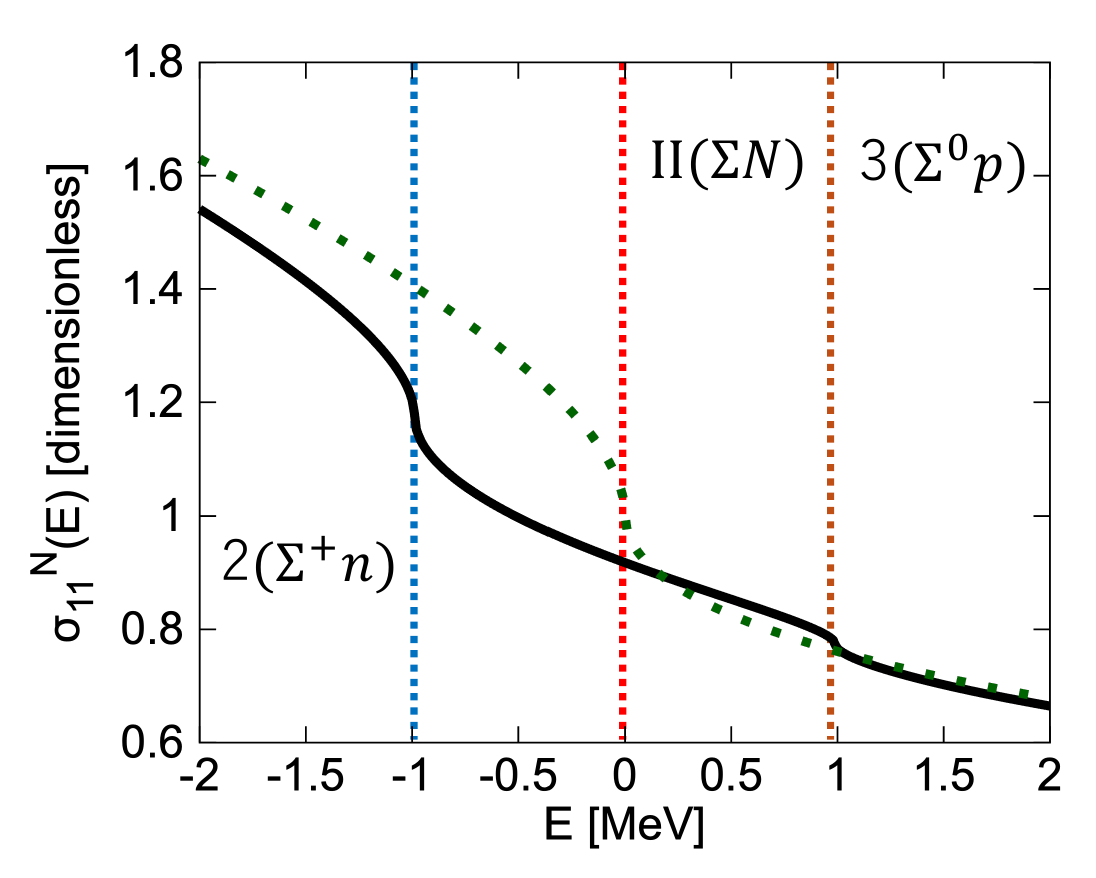}
    \caption{
    Same as Fig.~\ref{fig: F1}, but for $a_{1/2}=+1.0-i0.8~{\rm fm}$.
    }
    \label{fig: F2}
\end{figure}

We also consider the case with
\begin{align}
    a_{1/2} = +1.0 - i\,0.8~{\rm fm}, \quad (\Delta_{23}\to0).
    \label{eq: A +1}
\end{align}
For this value of $a_{1/2}$, a quasibound pole~\cite{Nishibuchi:2023acl} is generated below the $\Sigma N$ threshold. In this case, the corresponding $K$-matrix parameters are obtained as
\begin{align}
    C_1 &= -0.56~{\rm fm}, \quad
    C_2 = -1.1~{\rm fm}, \\
    C_3 &= -0.55~{\rm fm}, \quad
    C_4 = 0.55~{\rm fm},
\end{align}
and the scattering lengths in the isospin-broken case are
\begin{align}
    a_2 &= 0.68-i0.62~{\rm fm} \quad (\Delta_{23}\neq0),
    \label{eq: a2 F2}\\
    a_3 &= 0.26-i0.27~{\rm fm} \quad (\Delta_{23}\neq0).
    \label{eq: a3 F2}
\end{align}
The normalized cross section corresponding to $a_{1/2}$ in Eq.~\eqref{eq: A +1} is shown in Fig.~\ref{fig: F2}. In terms of the classification in Fig.~\ref{fig:four kinds}, both cusp structures correspond to type (c) reflecting the positive $\Re[a_{1/2}]$ in Eq.~\eqref{eq: A +1}. In the present example, the isospin-breaking effects are again small, as can be seen from Eqs.~\eqref{eq: a2 F2} and~\eqref{eq: a3 F2}, which approximately satisfy $a_2 \approx 2a_3$. Accordingly, the cusp structures at the $\Sigma^{+}n$ and $\Sigma^{0}p$ thresholds remain closely related. 

%%%%%%%%%%%%%%%%%%%%%%%%%%%%%%%%%%%%%%%
\subsection{
%\com{Chiral EFT result}
Numerical results with chiral EFT input
}\label{subsec: chiral EFT}
%%%%%%%%%%%%%%%%%%%%%%%%%%%%%%%%%%%%%%%

As a more realistic input, we consider the $\Sigma N$ analysis within chiral EFT at N$^2$LO~\cite{Haidenbauer:2023qhf}. In the present work, we do not aim at reproducing the full chiral-EFT analysis itself. Instead, we use our simplified $K$-matrix framework to examine how the cusp structures behave for a realistic set of low-energy inputs. Accordingly, the following discussion should be understood as an illustrative application of the present framework. Specifically, we use only the spin-triplet channel and ignore the spin-singlet component. In this setup, the analysis is restricted to the near-threshold $s$-wave amplitude. Hence, possible effects of higher partial waves and their couplings are not taken into account, while the importance of the $s$-$d$ mixing effects for the $\Sigma N$ cusp is discussed in Ref.~\cite{Haidenbauer:2021smk}.

\begin{figure}[tbp]
    \centering
    \includegraphics[width = 9cm, clip]{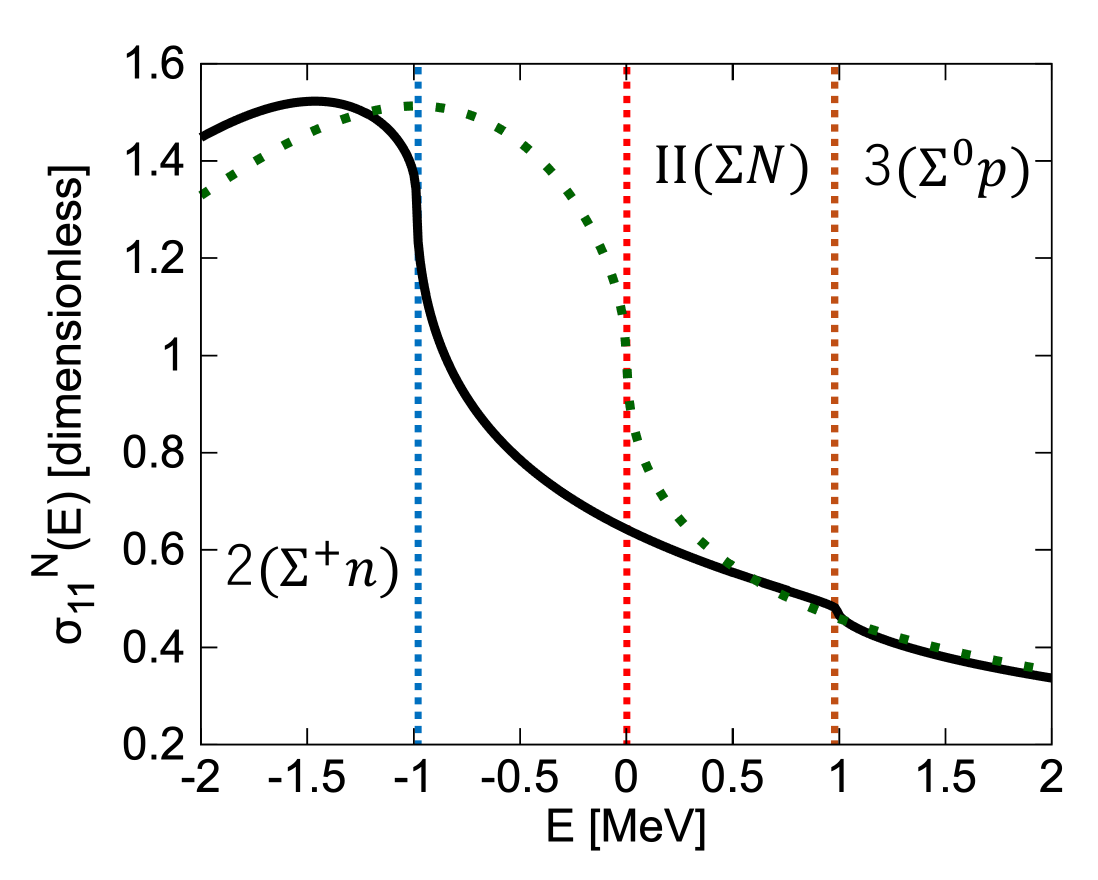}
    \caption{
    The normalized $\Lambda p$ elastic cross section for the chiral-EFT input with $C_1=-1.0~\mathrm{fm}$ and Eqs.~\eqref{eq: chiral1 C2}-~\eqref{eq: chiral1 C4}. The dotted and solid lines represent the cases with $\Delta_{23}\to0$ and $\Delta_{23}\neq0$, respectively. The vertical dotted lines indicate the thresholds of channels 2, II, and 3.
    }
    \label{fig: chiral1}
\end{figure}

The spin-triplet scattering lengths for the N$^2$LO model with cutoff $\Lambda=500$~MeV are given by~\cite{Haidenbauer:2023qhf}
\begin{align}
    a_{1/2} &= 2.60-i2.56~{\rm fm},
    \label{eq: a1/2 chiral} \\
    a_{3/2} &= 0.38~{\rm fm}.
    \label{eq: cp a3 H}
\end{align}
In the present analysis, these scattering lengths are used as input in the isospin-symmetric limit. Since the imaginary part of the $I=3/2$ scattering length vanishes identically in the isospin-symmetric case, Eqs.~\eqref{eq: a1/2 chiral} and~\eqref{eq: cp a3 H} provide only three independent constraints on the four parameters in Eq.~\eqref{eq: Kmatrix for lambdaN}. To determine all of them, we fix one parameter to be $C_1=-1.0~{\rm fm}$, which corresponds to a typical hadronic length scale. Then, the remaining parameters are uniquely determined from Eqs.~\eqref{eq: a1/2 chiral} and~\eqref{eq: cp a3 H} as
\begin{align}
    C_2 &= -4.29~{\rm fm}, \label{eq: chiral1 C2}\\
    C_3 &= -2.33~{\rm fm}, \\
    C_4 &= 1.35~{\rm fm}. \label{eq: chiral1 C4}
\end{align}
For this parameter set, the slopes of the $\Lambda p$ elastic cross section at the threshold of channel II in the isospin-symmetric limit are determined by the real and imaginary parts of
\begin{align}
    A-B=1.80-i2.56~{\rm fm}.
    \label{eq: A-B chiral1}
\end{align}
The slopes of the cross section associated with the degenerate 2nd and 3rd thresholds are given by
\begin{align}
    a_2 - b_{11}^{\rm 2nd} &=1.20-i1.71~{\rm fm} \quad (\Delta_{23}\to0), \\
    a_3 - b_{11}^{\rm 3rd} &= 0.60-i0.853~{\rm fm} \quad (\Delta_{23}\to0).
\end{align}
In the isospin-broken case $\Delta_{23}\neq0$, the slopes of the cross section at the thresholds of channels 2 and 3 are determined by
\begin{align}
    a_2 - b^{\rm 2nd}_{11} &= 0.898-i2.43~{\rm fm} \quad (\Delta_{23}\neq0), 
    \label{eq: slopes chiral1 2nd}\\
    a_3 - b^{\rm 3rd}_{11} &= 0.234-i0.670~{\rm fm} \quad (\Delta_{23}\neq0).
    \label{eq: slopes chiral1 3rd}
\end{align}
The corresponding normalized cross sections in the isospin-symmetric and isospin-broken cases are shown by the dotted and solid lines, respectively, in Fig.~\ref{fig: chiral1}.

The dotted line with $\Delta_{23}\to 0$ in Fig.~\ref{fig: chiral1} exhibits a cusp structure of type (c) at the threshold of channel II. The solid line with finite $\Delta_{23}$ also exhibits cusp structures of type (c) at both the 2nd ($\Sigma^+ n$) and 3rd ($\Sigma^0 p$) thresholds. Namely, the cusp structures are of the same type in both the isospin-symmetric and isospin-broken cases. These results imply that the isospin-breaking effects are not strong enough to change the types of cusps. However, 
Eqs.~\eqref{eq: slopes chiral1 2nd} and~\eqref{eq: slopes chiral1 3rd} show large deviations from the isospin relation $a_2-b^{\rm 2nd}_{11} = 2\left(a_3-b^{\rm 3rd}_{11}\right)$. Therefore, the isospin-breaking effects are substantial, even though the cusp structures remain of the same type in the present case.

%To see how the isospin-breaking effects modify the cusp structures, we compare the slopes of the cross sections in the isospin-broken and isospin-symmetric cases. In the isospin-symmetric limit, the slopes associated with the degenerate 2nd and 3rd thresholds are given by

%Comparing these values with Eqs.~\eqref{eq: slopes chiral1 2nd} and~\eqref{eq: slopes chiral1 3rd}, we find that the isospin-breaking effects do not change the signs of the real and imaginary parts of either $a_2 - b_{11}^{\rm 2nd}$ or $a_3 - b_{11}^{\rm 3rd}$, and hence the cusp type is unchanged. On the other hand, the magnitudes are modified significantly: the absolute value of the slope at the $\Sigma^+ n$ threshold becomes larger, so that the corresponding cusp is enhanced and becomes sharper, whereas that at the $\Sigma^0 p$ threshold becomes smaller, and the corresponding cusp is reduced. In this way, the isospin-breaking effects in the present example do not alter the cusp type itself, but they strongly modify the sharpness of the two cusp structures.

\begin{figure}[tbp]
    \centering
    \includegraphics[width = 9cm, clip]{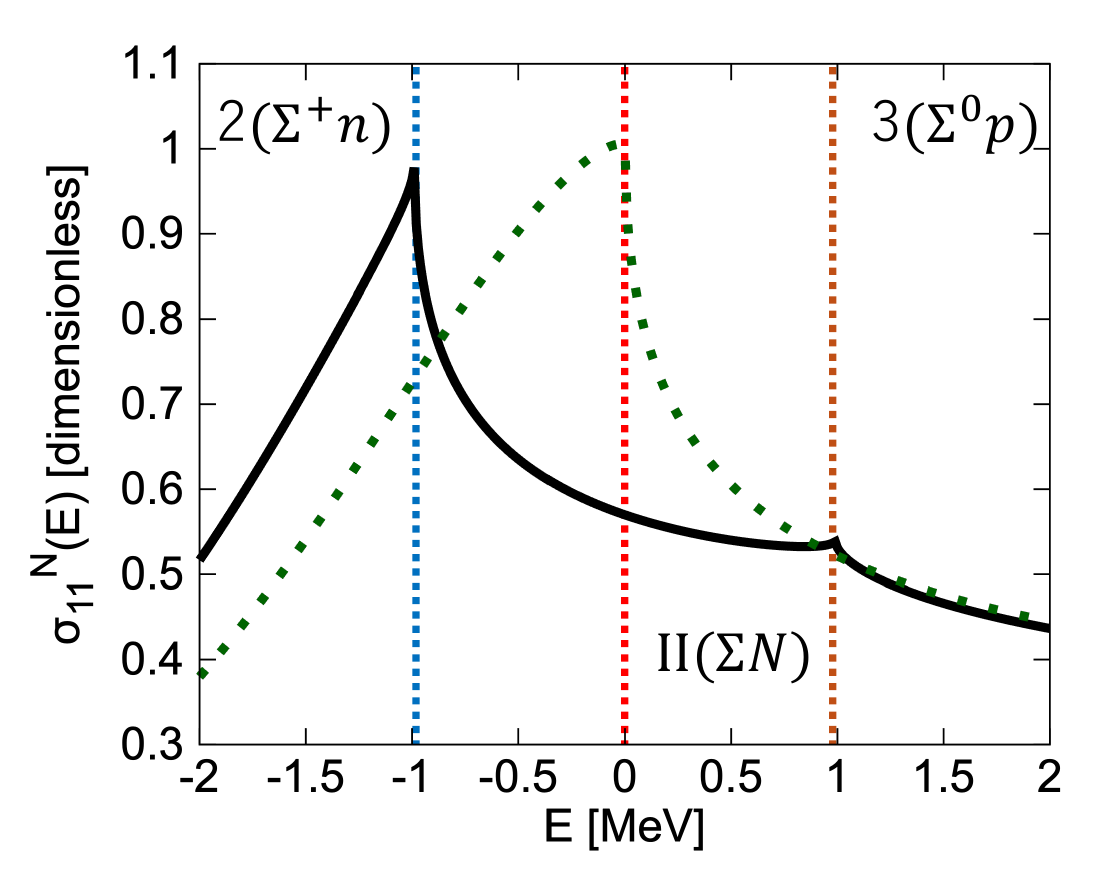}
    \caption{
    Same as Fig.~\ref{fig: chiral1}, but for $C_1=-10.0~{\rm fm}$ and Eqs.~\eqref{eq: chiral2 C2}-~\eqref{eq: chiral2 C4}.
    }
    \label{fig: chiral2}
\end{figure}

We next examine an extreme choice of the parameter $C_1$. For
$C_1=-10.0~{\rm fm}$ with fixed scattering lengths in Eqs.~\eqref{eq: a1/2 chiral} and~\eqref{eq: cp a3 H},
the remaining parameters are determined as
\begin{align}
    C_2 &= -26.2~{\rm fm}, \label{eq: chiral2 C2}\\
    C_3 &= -13.3~{\rm fm}, \\
    C_4 &= 11.0~{\rm fm}.\label{eq: chiral2 C4}
\end{align}
In this case, all parameters are much larger than the typical length scale in hadron physics. Nevertheless, the quantity determining the slopes of the cross section at the threshold of channel \ff is of the typical length scale:
\begin{align}
    A-B=0.180-i2.56~{\rm fm}.
    \label{eq: slopes II chiral2}
\end{align}
This is because the physical scattering lengths are fixed as in Eqs.~\eqref{eq: a1/2 chiral} and \eqref{eq: cp a3 H}. In other words, the large values of $C_{i}$ cancel each other, yielding a quantity of typical order in the form $A-B$.
The real part of $A-B$ has the opposite sign from that in Eq.~\eqref{eq: A-B chiral1}.  The slopes of the cross section associated with the degenerate 2nd and 3rd thresholds are given by
\begin{align}
    a_2 - b_{11}^{\rm 2nd} &= 0.120-i1.71~{\rm fm} \quad (\Delta_{23}\to0), 
    \label{eq: sym a2-b chiral2}\\
    a_3 - b_{11}^{\rm 3rd} &= 0.060-i0.853~{\rm fm} \quad (\Delta_{23}\to0).
    \label{eq: sym a3-b chiral2}
\end{align}
In the isospin-broken case $\Delta_{23}\neq0$, the slopes at the 2nd and 3rd thresholds are given by
\begin{align}
    a_2 - b^{\rm 2nd}_{11} &= -0.470-i2.43~{\rm fm} \quad (\Delta_{23}\neq0), 
    \label{eq: slopes chiral2 2nd}\\
    a_3 - b^{\rm 3rd}_{11} &= -0.231-i0.494~{\rm fm} \quad (\Delta_{23}\neq0).
    \label{eq: slopes chiral2 3rd}
\end{align}
In this case, the real parts in Eqs.~\eqref{eq: slopes chiral2 2nd} and~\eqref{eq: slopes chiral2 3rd} are changed from those in Eqs~\eqref{eq: sym a2-b chiral2} and~\eqref{eq: sym a3-b chiral2} in the isospin-symmetric limit. The corresponding normalized cross sections are shown in Fig.~\ref{fig: chiral2}.

Figure~\ref{fig: chiral2} shows that the dotted line, which corresponds to the isospin-symmetric case, exhibits a type-(c) cusp at the threshold of channel \ff,\footnote{Although it may not be clearly identified as type-(c) from the figure, this cusp is classified as type-(c) based on the sign of the real part of $A-B$ in Eq.~\eqref{eq: slopes II chiral2}, as discussed below.} whereas the solid line, which corresponds to the isospin-broken case, exhibits type-(a) cusp structures at the 2nd and 3rd thresholds. Thus, in the present example, the types of the cusp structures are changed by the isospin-breaking effects.

This is understood from the sign changes in the quantities determining the threshold slopes. In the isospin-symmetric limit, Eq.~\eqref{eq: slopes II chiral2} shows that the real part of $A-B$ is positive while its imaginary part is negative, which leads to the type-(c) cusp structure at channel \ff. On the other hand, in the isospin-broken case, Eqs.~\eqref{eq: slopes chiral2 2nd} and~\eqref{eq: slopes chiral2 3rd} show that both the real and imaginary parts are negative at the 2nd and 3rd thresholds, so that the cusp structures there are classified as type (a).

Comparing the isospin-symmetric result with the corresponding quantities in the isospin-broken case, we find that the real parts change their signs due to the isospin-breaking effects, while the imaginary parts remain negative. Therefore, the present example shows that the isospin-breaking effects can modify not only the sharpness of the cusp structures but also their type itself. However, it should be noted that realizing such large isospin-breaking effects requires  parameters with extreme values.

%resolution、実験との接続の話
%%%%%%%%%%%%%%%%%%%%%%%%%%%%%%%%%%%%%%%
\subsection{Effects of experimental resolution on cusp structures}
%%%%%%%%%%%%%%%%%%%%%%%%%%%%%%%%%%%%%%%

So far we have discussed the cusp structures at the exact threshold energies. In an actual
measurement, however, the observed line shape is smeared by the finite energy resolution of the
detector, and the available statistics are finite. Here we illustrate, using the chiral-EFT
examples of Sec.~\ref{subsec: chiral EFT}, how the isospin-breaking effects discussed above appear
under realistic experimental conditions. We consider the $\Sigma N$ cusp measurement via the
$d(K^-,\pi^-)$ reaction proposed as the J-PARC E90 experiment, which is designed to achieve an
energy resolution of $\sigma_{\rm res}=0.4$~MeV~\cite{E90proposal}. Combined with a
neural-network-based background suppression technique, about $3\times10^4$ signal events are
expected to be usable for the line-shape analysis~\cite{Amemiya:2026ptep}. We also consider a
possible future improvement at the HIHR (high intensity and high resolution) beamline planned for the extended J-PARC hadron facility,
which is expected to reach $\sigma_{\rm res}=0.17$~MeV ($0.4$~MeV in FWHM)~\cite{Aoki:2021hihr}.

\begin{figure}[tbp]
    \centering
    \includegraphics[width = 9cm, clip]{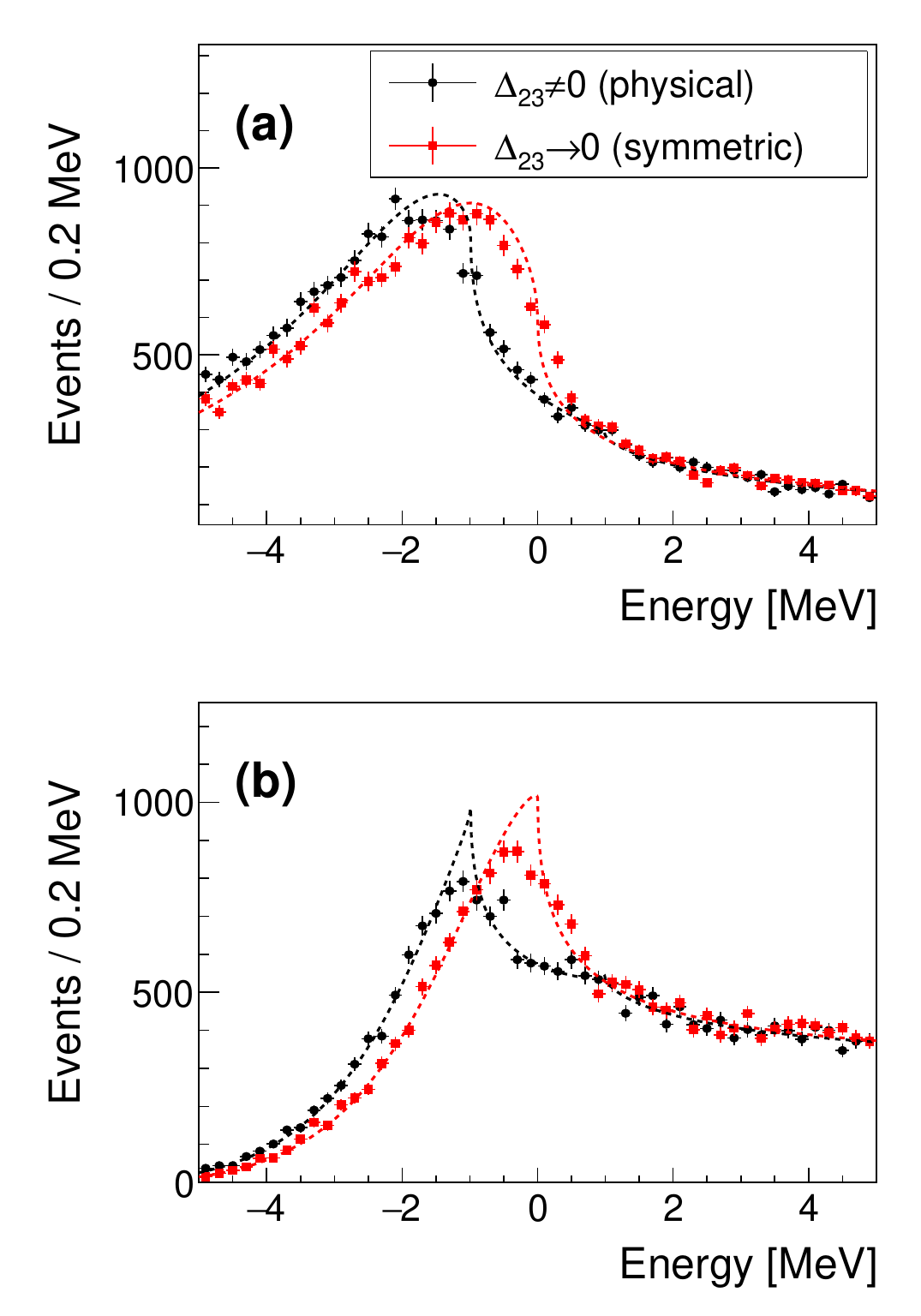}
    \caption{Physical ($\Delta_{23}\neq0$) vs.\ isospin-symmetric ($\Delta_{23}\to0$) cross
    sections, smeared with the J-PARC E90 resolution $\sigma_{\rm res}=0.4$~MeV for $3\times10^4$ events
    (points; legend applies to both panels), for (a) $C_1=-1.0$~fm and (b) $C_1=-10.0$~fm.
    Dashed curves: no resolution smearing (arbitrarily normalized for comparison).}
    \label{fig: referee res combined}
\end{figure}

Figure~\ref{fig: referee res combined} shows, for (a) $C_1=-1.0$~fm and (b) $C_1=-10.0$~fm, the physical ($\Delta_{23}\neq0$) and isospin-symmetric ($\Delta_{23}\to0$) cross sections after
smearing with the E90 resolution and statistics. A difference of this size remains visible after
smearing; the future HIHR resolution would make it even clearer.

\begin{figure}[tbp]
    \centering
    \includegraphics[width = 9cm, clip]{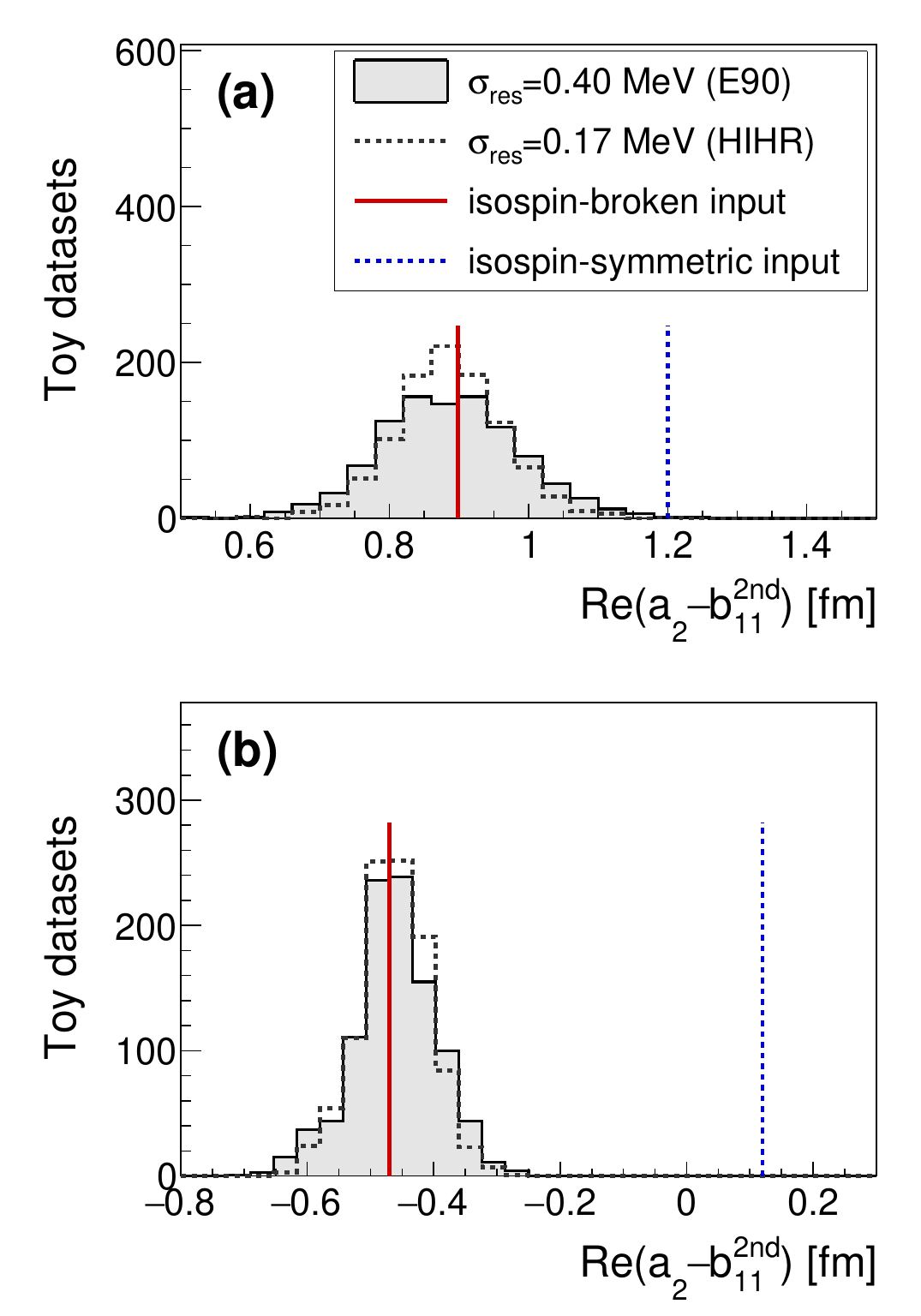}
    \caption{Distribution of the fitted $\mathrm{Re}[a_2-b_{11}^{\rm 2nd}]$ over 1000 independent
    pseudo-data sets, for (a) $C_1=-1.0$~fm and (b) $C_1=-10.0$~fm, comparing
    $\sigma_{\rm res}=0.4$~MeV (E90, shaded) and $0.17$~MeV (HIHR,
    dashed), both with $3\times10^4$ events. Vertical lines
    show the input isospin-broken (solid red, $\Delta_{23}\neq0$) and isospin-symmetric-limit
    (dashed blue, $\Delta_{23}\to0$) values used to generate the pseudo-data: $0.898$ and
    $1.20$~fm for $C_1=-1.0$~fm, $-0.470$ and $+0.120$~fm for $C_1=-10.0$~fm.}
    \label{fig: referee slope hist}
\end{figure}

For $C_1=-10.0$~fm, isospin breaking flips the sign of the near-threshold slope, changing the
cusp type from a rounded step ($\mathrm{Re}[a_2-b_{11}^{\rm 2nd}]=+0.12$~fm, isospin-symmetric
limit) to a genuine cusp ($-0.47$~fm, physical case). To quantify whether this sign change can be resolved under realistic experimental conditions, we generate 1000 independent pseudo-data sets and fit each of them to estimate the distribution of $\mathrm{Re}[a_2-b_{11}^{\rm 2nd}]$.
Figure~\ref{fig: referee slope hist} shows that fits to pseudo-data recover this sign at both the
E90 and HIHR resolutions, with a comparable statistical spread of the fitted
$\mathrm{Re}[a_2-b_{11}^{\rm 2nd}]$ (RMS $\simeq0.07$~fm at E90 and $0.06$~fm at HIHR), so the cusp
type remains distinguishable in practice. The $C_1=-1.0$~fm case, shown for comparison, exhibits no
such sign change, with a similar RMS spread ($\simeq0.10$~fm at E90 and $0.08$~fm at HIHR).

Altogether, these examples show that the resolution and statistics of J-PARC E90, and
prospectively of HIHR, are sufficient not only to reveal the isospin-breaking effects discussed
above in the observed line shape, but also to determine the cusp type quantitatively from a fit,
giving a concrete experimental handle on the phenomena analyzed in this work.

%%%%%%%%%%%%%%%%%%%%%%%%%%%%%%%%%%%%%%%
\section{Summary}
%%%%%%%%%%%%%%%%%%%%%%%%%%%%%%%%%%%%%%%

In this paper, we study the isospin-breaking effects on threshold cusp structures in the multichannel scattering and discuss the cusp structures in the $\Lambda N$-$\Sigma N$ system as a concrete example. In Sec.~\ref{subsec: N channel}, using the $K$-matrix, we derive a general expression for the near-threshold scattering amplitude in the $N$-channel case. With this expression, the slopes of the cross section at the threshold can be written in a transparent form in terms of the scattering length $a_N$ and the complex constants $b_{ij}$ appearing in the numerator of the scattering amplitude, which is suitable for discussing cusp structures.

We then classify the four possible cusp structures by the signs of the slopes above and below the threshold in Sec.~\ref{eq: cusp}, as shown in Fig.~\ref{fig:four kinds}. We also show that additional constraints on the cusp structures appear at the second- and third-lowest thresholds. In two-channel scattering, the cusp structure of the $(1,1)$ component of the cross section at the channel 2 threshold is restricted to the types (a) and (c) in Fig.~\ref{fig:four kinds} due to the absence of the imaginary part of $b_{11}$. We then discuss three-channel scattering and show that a similar restriction appears for the $(1,2)$ and $(2,1)$ components at the third-lowest threshold due to the absence of the imaginary parts of $b_{12}$ and $b_{21}$.

We also discuss the threshold cusp behavior of the Flatt\'e amplitude. While the near-threshold properties of the Flatt\'e form have been discussed in Ref.~\cite{Baru:2004xg}, we revisit them in our formulation based on the $K$-matrix representation. As a result, we show that the complex constants $b_{ij}$ do not appear in any component of the Flatt\'e amplitude, and hence the possible cusp structures are restricted to the types (a) and (c) in Fig.~\ref{fig:four kinds}.

Conversely, this implies that, for thresholds exhibiting type-(b) and type-(d) cusps, there exist sufficiently large values of $b_{ij}$ that can reverse the sign of the imaginary part of the scattering length. This indicates that the contribution of $b_{ij}$, or equivalently that of the associated background amplitude, plays an important role. 
Even when type-(a) and type-(c) cusps appear, however, the extent to which the contribution of $b_{ij}$ is involved cannot be determined a priori. Therefore, in general, the determination of the coefficient $b_{ij}$ is essential for extracting the scattering length from the cusp structure.

We then discuss the case of three-channel scattering with two nearby thresholds in Sec.~\ref{subsec : close threshold}. When the effects of the energy difference $\Delta_{23}$ between the two thresholds are sufficiently small, the cusp structures at the two thresholds are closely related, because their leading contributions have the same signs and therefore generate cusp structures of the same type. In the $\Delta_{23}\to0$ limit, the two cusps merge into a single cusp at the degenerate threshold, and the slope of this cusp is given by the sum of the corresponding leading contributions at the two separate thresholds.

In Sec.~\ref{sec : LN-SN scattering}, we discuss the isospin-breaking effects on threshold cusp structures in the $\Lambda p$-$\Sigma^{+}n$-$\Sigma^{0}p$ system with charge $Q=+1$ using the isospin-symmetric $K$-matrix. In Sec.~\ref{subsec: cusp lambda p}, we show that, when the isospin-breaking effects are small, isospin symmetry constrains not only the signs of the slopes of the cross section at the $\Sigma^+ n$ and $\Sigma^0 p$ thresholds but also their ratio. In the $\Lambda N$-$\Sigma N$ system with charge $Q=+1$, the leading contributions to the slope at the $\Sigma^+ n$ is twice that at the $\Sigma^0 p$ threshold. Therefore, when the isospin-breaking effects are small, the two cusp structures at the $\Sigma^+ n$ and $\Sigma^0 p$ thresholds are expected to be of the same type, with the cusp at the $\Sigma^+ n$ threshold being more prominent than that at the $\Sigma^0 p$ threshold.

In Sec.~\ref{sec: cs fg}, we perform the numerical calculations for parameter sets satisfying the separable condition of the $K$-matrix where the amplitude corresponds to the Flatt\'{e} amplitude. In these examples, the isospin-breaking effects are small, and Figs.~\ref{fig: F1} and~\ref{fig: F2} show that the cusp structures at the $\Sigma^+ n$ and $\Sigma^0 p$ thresholds remain of the same type, consistently with the general discussion. At the same time, the cusp at the $\Sigma^+ n$ threshold is more prominent than that at the $\Sigma^0 p$ threshold, reflecting the isospin relation discussed in Sec.~\ref{subsec: cusp lambda p}.

In Sec.~\ref{subsec: chiral EFT}, we perform numerical calculations using the N$^2$LO chiral-EFT input for the spin-triplet channel. In these examples, the isospin-breaking effects can be sizable. As shown in Fig.~\ref{fig: chiral1}, one parameter set leads to cusp structures of the same type, while their relative sharpness is strongly modified. By contrast, Fig.~\ref{fig: chiral2} shows that, for another parameter set with values an order of magnitude larger than typically expected, the cusp type itself changes, because the isospin-breaking effects modify the signs of slopes of the cross section. These results show that, in the $\Lambda N$-$\Sigma N$ system, isospin breaking can affect not only the relative strength of the cusp structures but also their qualitative shape.

The present study provides a general framework for analyzing threshold cusp structures in systems with nearby thresholds and clarifies how isospin symmetry and its breaking constrain the cusp behavior. This viewpoint should be useful for future analyses of threshold phenomena in hadron scattering and near-threshold states. For instance, the $T_{cc}$ state in the $D^{0}D^{*+}$-$D^{*0}D^{+}$ system, the $f_{0}(980)$ and $a_{0}(980)$ states in the $K^{+}K^{-}$-$K^{0}\bar{K}^{0}$ system, and the $\Lambda(1405)$ state in the $K^{-}p$-$\bar{K}^{0}n$ system are typical examples of states with nearly degenerate thresholds. The framework developed here thus provides a useful basis for a unified understanding of cusp phenomena in near-threshold hadronic systems.

\begin{acknowledgments}
The authors are grateful to Christoph Hanhart, Johann Haidenbauer, and Kiyoshi Tanida for useful discussions.
This work has been supported in part by 
JSPS KAKENHI Grant Numbers
JP26K07088, % Kiban C (Hyodo)
JP26H01426, % Gakuhen A Koubo (Hyodo)
JP25KJ1996,  %Gakushin (Sone)
JP25H01272, %Gakuhen A (Ichikawa)
JP23H05439, % Kiban S (Hyodo)
 and 
JP22K03637, % Kiban C (Hyodo)
%put your acknowledgments here.
by the RCNP Collaboration Research network (COREnet) 048 "Revealing the nature of exotic hadrons in Belle (II) by collaboration of experimentalists and theorists", 
and by 
 %Gakushin
MIYAKO-MIRAI Project of Tokyo Metropolitan University. 
% MIYAKO MIRAI PROJECT (Sone)

\end{acknowledgments}

\appendix
%
%%%%%%%%%%%%%%%%%%%%%%%%%%%%%%%%%%%%

\section{Isospin-basis scattering amplitude with a separable $K$-matrix}
\label{app:separable_K_isospin}

In this Appendix, we discuss the implications of the separable $K$-matrix condition for the $\Lambda N$-$\Sigma N$ scattering amplitude in the isospin basis, presented in Sec.~\ref{sec: cs fg}. In particular, we show that a separable $K$-matrix cannot describe nonzero interactions in both the $I=1/2$ and $I=3/2$ $\Sigma N$ channels simultaneously. Throughout this Appendix, we consider the isospin-symmetric limit $\Delta_{23}\to0$.

We first introduce the relation between the charge basis and the isospin basis. We focus on the $I_3=1/2$ sector and define
\begin{align}
    \ket{I=1/2;\Lambda N}
    &=
    \ket{\Lambda p},
    \label{eq: LambdaN isospin}
    \\
    \ket{I=1/2;\Sigma N}
    &=
    \sqrt{\frac{2}{3}}\ket{\Sigma^+n}
    +
    \sqrt{\frac{1}{3}}\ket{\Sigma^0p},
    \label{eq: SigmaN isospin 1/2}
    \\
    \ket{I=3/2;\Sigma N}
    &=
    -
    \sqrt{\frac{1}{3}}\ket{\Sigma^+n}
    +
    \sqrt{\frac{2}{3}}\ket{\Sigma^0p}.
    \label{eq: SigmaN isospin 3/2}
\end{align}
The transformation matrix from the charge basis to the isospin basis is therefore given by
\begin{align}
    \hat{U}
    &=
    \begin{pmatrix}
        1 & 0 & 0\\
        0 & \sqrt{2/3} & \sqrt{1/3}\\
        0 & -\sqrt{1/3} & \sqrt{2/3}
    \end{pmatrix},
    \label{eq: U for isospin} \\
    \begin{pmatrix}
        \ket{I=1/2;\Lambda N} \\
        \ket{I=1/2;\Sigma N} \\
        \ket{I=3/2;\Sigma N}
    \end{pmatrix}
    &=
    \hat{U}
    \begin{pmatrix}
        \ket{\Lambda p} \\
        \ket{\Sigma^+n} \\
        \ket{\Sigma^0p}
    \end{pmatrix}
\end{align}
Using this unitary transformation, the $K$-matrix in the charge basis in Eq.~\eqref{eq: Kmatrix for lambdaN} is transformed into the isospin basis as
\begin{align}
    \hat{K}_{I}
    &=
    \hat{U}\hat{K}\hat{U}^{\dagger}
    \notag \\
    &=
    \begin{pmatrix}
        C_1 & \sqrt{3}C_4 & 0 \\
        \sqrt{3}C_4 & 2C_2-C_3 & 0\\
        0 & 0 & -C_2+2C_3
    \end{pmatrix}.
    \label{eq: Kmat isospin}
\end{align}
The upper-left $2\times2$ block represents the coupled $\Lambda N$-$\Sigma N$ interactions in the $I=1/2$ sector, while the $(3,3)$ component represents the $\Sigma N$ interaction in the $I=3/2$ sector. The momentum matrix in Eq.~\eqref{eq: pmat del0} is invariant under the transformation to the isospin basis:
\begin{align}
    \hat{p}_{I}(E)
    =
    \hat{U}\hat{p}(E)\hat{U}^{\dagger}
    =
    \begin{pmatrix}
        p_1 & 0 & 0\\
        0 & p_{\rm \ff}(E) & 0\\
        0 & 0 & p_{\rm \ff}(E)
    \end{pmatrix}
    =
    \hat{p}(E).
    \label{eq: pmat isospin App}
\end{align}

Substituting Eqs.~\eqref{eq: Kmat isospin} and~\eqref{eq: pmat isospin App} into Eq.~\eqref{eq: f^N2}, we find that the $\Sigma N$ elastic amplitude in the $I=1/2$ channel can be written in the effective-range form near the $\Sigma N$ threshold as
\begin{align}
    F_{1/2}^{\Sigma N}(E)
    &=
    \frac{1}{-\frac{1}{a_{1/2}}-ip_{\rm \ff}(E)},
    \label{eq: F1/2 Sigma N}\\
    a_{1/2}
    &=
    -\frac{
        2C_2-C_3
        -i\qty{C_1(2C_2-C_3)-3C_4^2}p_1
    }{
        1-iC_1p_1
    }.
    \label{eq: a1/2 App}
\end{align}
Similarly, the $\Sigma N$ elastic amplitude in the $I=3/2$ channel can be written in the effective-range form as
\begin{align}
    F_{3/2}^{\Sigma N}(E)
    &=
    \frac{1}{-\frac{1}{a_{3/2}}-ip_{\rm \ff}(E)},
    \label{eq: F3/2 Sigma N}
    \\
    a_{3/2}
    &=
    C_2-2C_3.
    \label{eq: a3/2 App}
\end{align}
These scattering lengths satisfy the relations in Eqs.~\eqref{eq: a2/2 a2 a3} and~\eqref{eq: a3/2 a2 a3}.

We now impose the separable condition on the $K$-matrix. The Flatt\'e amplitude discussed in Sec.~\ref{subsec: Flatte} is obtained from the separable $K$-matrix in Eq.~\eqref{eq: K matrix for Flatte}. A nonzero separable matrix has rank one. Therefore, for the $3\times3$ $K$-matrix in Eq.~\eqref{eq: Kmatrix for lambdaN}, the separable condition is equivalent to the vanishing of all $2\times2$ minor determinants. An equivalent set of conditions is
\begin{align}
    (C_2-2C_3)(2C_2-C_3)&=0,
    \label{eq: branch condition}
    \\
    C_4(C_2-2C_3)&=0,
    \label{eq: App condition F1}
    \\
    C_1C_3-C_4^2&=0,
    \label{eq: App condition F2}
    \\
    C_1C_2-2C_4^2&=0.
    \label{eq: 4th condition}
\end{align}
The first condition, Eq.~\eqref{eq: branch condition}, shows that the solutions are divided into two branches, $C_2=2C_3$ and $2C_2=C_3$.

First, let us consider the branch $C_2=2C_3$. In this branch, Eq.~\eqref{eq: App condition F1} is automatically satisfied, and Eqs.~\eqref{eq: App condition F2} and~\eqref{eq: 4th condition} reduce to the single condition
\begin{align}
    C_1C_3-C_4^2=0.
    \label{eq: branch1 rank1 condition}
\end{align}
Using $C_4=\pm\sqrt{C_1C_3}$, the isospin-basis $K$-matrix can be expressed in terms of $C_1$ and $C_3$, up to the sign of $C_4$, as
\begin{align}
    \hat{K}_{I}
    &=
    \begin{pmatrix}
        C_1 & \pm\sqrt{3C_1C_3} & 0 \\
        \pm\sqrt{3C_1C_3} & 3C_3 & 0\\
        0 & 0 & 0
    \end{pmatrix}.
    \label{eq: KI under 1}
\end{align}
From Eq.~\eqref{eq: KI under 1}, one can see that the $I=3/2$ interaction vanishes, while the $I=1/2$ block can remain nonzero. In this case,
\begin{align}
    a_{3/2}=0,
    \label{eq: a32 zero branch1}
\end{align}
whereas $a_{1/2}$ in Eq.~\eqref{eq: a1/2 App} can be nonzero. This is the branch used in the numerical examples in Sec.~\ref{sec: cs fg}. Note that the choice of the sign of $C_{4}$ does not affect the result, since the sign of the off-diagonal components is irrelevant.

Next, we consider the other branch, $2C_2=C_3$. In this branch, Eq.~\eqref{eq: App condition F1} becomes
\begin{align}
    -\frac{3}{2}C_3C_4=0.
\end{align}
For nonzero $C_2$ and $C_3$, this condition requires $C_4=0$. The remaining conditions, Eqs.~\eqref{eq: App condition F2} and~\eqref{eq: 4th condition}, then reduce to the same condition,
\begin{align}
    C_1C_3=0,
\end{align}
which gives $C_{1}=0$.
Thus, for nonzero $C_3$, one obtains
\begin{align}
    C_1=0,
    \qquad
    C_4=0,
    \qquad
    C_3=2C_2.
    \label{eq: branch2 condition}
\end{align}
The corresponding isospin-basis $K$-matrix then takes the form
\begin{align}
    \hat{K}_{I}
    =
    \begin{pmatrix}
        0 & 0 & 0 \\
        0 & 0 & 0 \\
        0 & 0 & 3C_3/2
    \end{pmatrix}.
    \label{eq: KI under 2}
\end{align}
In this branch, all interactions in the $I=1/2$ sector vanish, while the $I=3/2$ interaction remains nonzero. Therefore, the scattering length $a_{1/2}$ vanishes, whereas
\begin{align}
    a_{3/2}=-\frac{3}{2}C_3
\end{align}
is nonzero. Because $C_4=0$, this branch does not generate the $\Lambda N$-$\Sigma N$ coupling, and hence no cusp structure appears in the $\Lambda p$ elastic cross section at the $\Sigma N$ threshold. In this way, when the $K$-matrix in the charge basis is separable, the interactions in the $I=1/2$ and $I=3/2$ sectors cannot both remain nonzero simultaneously.\footnote{More precisely, there exist trivial solutions: $C_{3}=0$ in Eq.~\eqref{eq: KI under 1} (interaction only in the $\Lambda N$ channel), $C_{1}=0$ in Eq.~\eqref{eq: KI under 1} (interaction only in the $I=1/2$, $\Sigma N$ channel), and $C_{1}=C_{3}=0$ in Eq.~\eqref{eq: KI under 1}, or $C_{3}=0$ in Eq.~\eqref{eq: KI under 2} (no interaction at all). }

The above result can also be understood directly from the rank of the $K$-matrix. A unitary transformation does not change the rank of a matrix:
\begin{align}
    {\rm rank}(\hat{K}_{I})
    =
    {\rm rank}(\hat{U}\hat{K}\hat{U}^{\dagger})
    =
    {\rm rank}(\hat{K}).
    \label{eq: rank invariant App}
\end{align}
On the other hand, in the isospin basis, the $K$-matrix in Eq.~\eqref{eq: Kmat isospin} is block diagonal. If both the $I=1/2$ block and the $I=3/2$ block are nonzero, the rank of $\hat{K}_{I}$ is at least two. Therefore, a rank-one separable $K$-matrix cannot have nonzero interactions in both isospin sectors simultaneously. One of the two isospin-sector interactions must vanish. Consequently, the separable Flatt\'e-type $K$-matrix can describe either a nonzero $I=1/2$ interaction with $a_{3/2}=0$ or a nonzero $I=3/2$ interaction with $a_{1/2}=0$, but not both simultaneously.
Physically, this can be understood as follows: the bare state with mass $E_{\rm BW}$ introduced in Eq.~\eqref{eq: Flatte for N} carries a definite isospin, and therefore can represent only either the $I=1/2$ or the $I=3/2$ interaction.

%Create the reference section using BibTeX:
%\bibliographystyle{h-physrev3}
%\bibliography{Sone2026jpr}

\end{document}